# Epidemic Informatics and Control: A Holistic Approach from System Informatics to Epidemic Response and Risk Management in Public Health

Hui Yang[1*], Siqi Zhang[1], Runsang Liu[1], Alexander Krall[1], Yidan Wang[1], Marta Ventura[1], and Chris Deflitch[2]

[1] Center for Health Organization Transformation
The Pennsylvania State University, University Park, PA, USA

[2]Department of Emergency Medicine
Vice president, Penn State Health
Chief medical information officer, Milton S. Hershey Medical Center

## Abstract

Epidemic outbreaks such as Coronavirus disease 2019 (COVID-19) impact the health of our society and bring significant disruptions to the US and the world. Each country has to dynamically adjust health policies, plan healthcare resources, control travels with little time latency to mitigate risks and safeguard the population. With rapid advances in health and information technology, more and more data are collected in the dynamically evolving process of epidemic outbreaks. The availability of data calls upon the development of analytical methods and tools to gain a better understanding of virus spreading dynamics, optimize the design of healthcare policies for epidemic control, and improve the resilience of health systems. This paper presents a holistic systems informatics approach, i.e., **D**efine**, M**easure, **A**nalyze, **I**mprove, and **C**ontrol (DMAIC), for epidemic response and management through the intensive use of data, statistics and optimization. Despite the sustained successes of system informatics in a variety of established industries such as manufacturing, logistics, services and beyond, there is a dearth of concentrated review and application of the data-driven DMAIC approach in the context of epidemic outbreaks. First, we define specific challenges posed by epidemic outbreaks to populational health, health systems, as well as economic challenges to different industries such as retailing, education and manufacturing. Second, we present a review of medical testing and statistical sampling methods for data collection, as well as existing efforts in data management and data visualization. Third, we discuss the importance to realizing the full potential of data for epidemic insights, and emphasize the need to leverage analytical methods and tools for decision support. Fourth, an epidemic brings imperative changes to health systems. We discuss the new trend of healthcare solutions to improve system resilience, including telehealth, artificial intelligence, resource allocation, and system re-design. In closing, prescriptive approaches are discussed to optimize the health policies and action strategies for controlling the spread of virus. We posit that this work will catalyze more in-depth investigations and multi-disciplinary research efforts to accelerate the application of system informatics methods and tools in epidemic response and risk management.

**Keywords**: Infectious disease, epidemic, diagnostic testing, risk management, health systems, data analytics, simulation modeling, artificial intelligence

# I. Introduction

Epidemic outbreaks impact the health of our society and bring significant disruptions to the US and the world. For example, Coronavirus Disease 2019 (COVID-19) is currently ravaging multiple countries and was declared as a global pandemic by the World Health Organization (WHO) in March 2020. COVID-19 has caused a total of approximately 7.82 million infected cases and 432K deaths worldwide, as well as 2.17 million infected cases and 118K deaths in the US by June 16, 2020 [1]. The abrupt increase of cases quickly exceeds the capacity of health systems and highlights the shortages of workers, beds, medical supplies and equipment. Many governments have taken a variety of actions (e.g., lockdown, large-scale testing, stay-at-home) to flatten the curve and avoid overwhelming health systems, but these reactionary policies have resulted in great economic losses. The US unemployment rate has skyrocketed from 3.5% in February 2020 to 14.7% in April 2020 [2]. The number of unemployed persons has increased to 23.1 million, which is even worse than the Great Depression in 1930s. The economic uncertainty has caused US stock markets to trigger the circuit breakers to halt trading for a historical 4 times in the week of March 9-16, 2020 [3]. The US GDP shrunk 4.8% in the first quarter of 2020.

When the COVID-19 epidemic emerged, it was not uncommon to encounter a misperception or misinformation that coronavirus is like the seasonal influenza (flu). Although there are similarities (e.g., causing respiratory illness) between coronavirus and flu virus, they are significantly different. COVID-19 or severe acute respiratory syndrome (SARS) is caused by the family of coronavirus, which is not the same as the flu virus. There are three major types of flu viruses - Types A, B and C. Type A flu virus caused many epidemics in the past 100 years (e.g., 1918 Spanish Flu [4], 1968 H3N2 epidemic [5], and 2009 H1N1 epidemic [6]). It is worth mentioning that Type A flu virus infects a wide variety of animals (e.g., poultry, swine, aquatic birds) and easily evolves and mutates genes. Once transported and adapted to humans, it can evolve into an epidemic. Types B and C flu viruses infect only humans as the typical seasonal flu and has rarely been the cause of past epidemics [7]. It is estimated by Center for Disease Control and Prevention (CDC) that seasonal flu causes approximately 140,000 – 810,000 hospitalizations and 12,000 – 61,000 deaths annually since 2010 [8]. However, the death toll of 1918 Spanish Flu is about 50 million worldwide and 675,000 in the US.

Historically, epidemics are inevitable and recur at more or less near-periodic cycles. It is difficult to predict when a new virus will emerge and cause an epidemic. The infection rate of a virus is commonly measured by the basic reproduction number $R_0$, which characterizes how many people on average can be infected by one infected individual in a susceptible population. For COVID-19, $R_0$ is estimated to range from 1.4 to 6.49, with a mean of 3.28 [9]. The potential transmission pathway can be either through air droplets, which are generated when infected individuals talk, cough, or sneeze, or through contact with an infected person or surface that is contaminated with the virus. At the start of an outbreak, antivirals and vaccines are often not available. People can only resort to non-pharmaceutical interventions (NPIs) for the control and containment of virus spread [10]. Traditional NPI methods include the practice of good personal hygiene, the use of disinfectants, the isolation and quarantine of infected individuals, and the limitation of public gatherings. From 1918 Spanish flu epidemic to COVID-19, this situation does not change much although health systems become more advanced and medical resources are richer than before.

However, one thing that does change is the faster and augmented capability of medical testing and diagnostics, thanks to rapid advances of gene/DNA, microbiology, and imaging technologies [11]. As such, large amounts of data are collected in the evolving process of epidemic outbreaks. The availability of data calls upon the development of analytical methods and tools to gain a better understanding of virus spreading dynamics, optimize the design of healthcare policies for epidemic control, and improve the resilience of health systems. Therefore, this paper presents a holistic system informatics approach of **D**efine**, M**easure, **A**nalyze, **I**mprove, and **C**ontrol (DMAIC) for epidemic response and management through the intensive use of data, statistics and optimization. Despite the sustained successes of DMAIC in a variety of established industries such as manufacturing, logistics, services and beyond [12-14], there is a dearth of concentrated review and application of the data-driven DMAIC approach in the context of epidemic outbreaks. As shown in Figure 1, The DMAIC methodology consists of five phases: **1) Define**: outline the societal challenges posed by the epidemic; **2) Measure**: collect data about key variables in the epidemic process; **3) Analyze**: extract useful information pertinent to the spread of epidemic; **4) Improve**: design solutions and methods to improve the resilience of health systems; **5) Control**: develop health policies, management plans, and intervention methods to control the spread of infectious diseases. The goal of this paper is to catalyze more in-depth investigations and multi-disciplinary research efforts to accelerate the application of system informatics methods and tools in epidemic response and risk management.

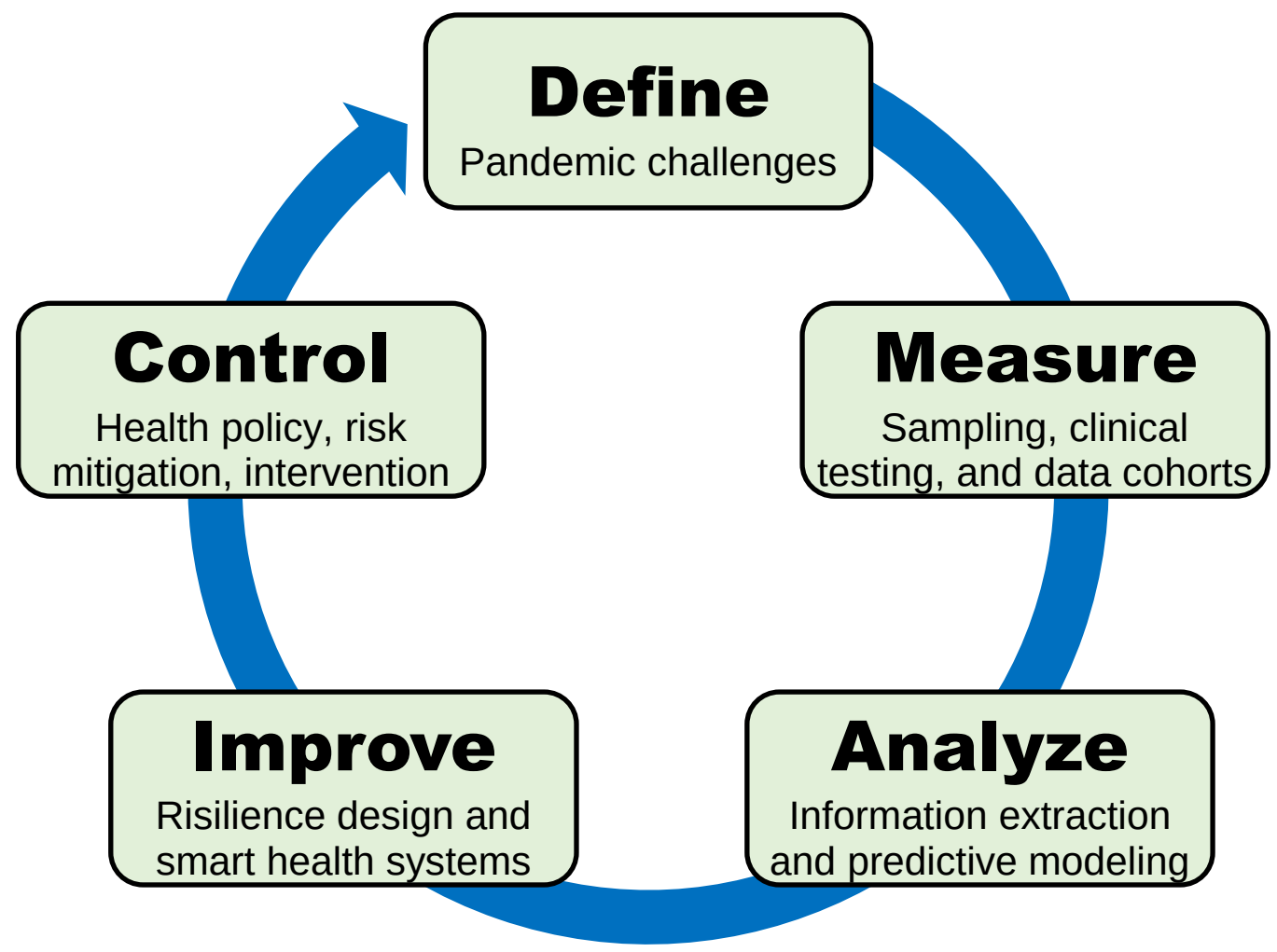

**Figure 1. The flowchart of system informatics for epidemic response and risk management**

The remainder of this paper is organized as follows: Section II discusses specific societal challenges arising from large-scale outbreaks of infectious diseases. Section III reviews the sampling and testing strategies to increase information visibility for epidemic management. Then, we present a review of analytical methods and tools for the extraction of useful information in Section IV. Continuous improvements and re-design to improve the resilience of health systems are discussed in Section V, and Section VI presents the health policies and intervention strategies for the control of virus spread. Section VII discusses the system informatics approach for epidemic management and concludes this paper.

# II. Epidemic Challenges to Our Society

## A. Health System Challenges

Epidemic outbreak calls upon the execution of large amounts of clinical testing to examine the prevalence of a virus in the population. No doubt, such a large demand poses significant challenges on the manufacturing and supply chain systems. Fortunately, advanced medical technology (e.g., gene/DNA, microbiology) enables the provision of viral and/or antibody testing kits to the US population. For example, as of June 19, 2020, there are a total of 26,781,666 viral tests performed to determine whether an individual is currently infected by the coronavirus [1]. Approximately 10% of the test results are positive. Among a sample of 1,934,566 individuals with COVID-19, most of them are within 18-44 and 45-64 age groups (41.4% and 32.8%, respectively). For the rest, 5.1% and 9.5% are aged 0-17 and 65-74, respectively, and 11% of them are above 75 [1]. In general, when the age of patients increases, the hospitalization rate also becomes higher. Hospitalization rate is the ratio between the number of individuals who are hospitalized within 14 days after a positive viral test and the total population in a spatial region. As shown in Table 1, the overall cumulative hospitalization rate is 94.5 per million [1]. For people aged 50-64 and above 65, the rates increase to 143 and 286.9 per million, respectively. However, for people aged 0-4 and 5-17, the rates declined to 7.4 and 3.5, respectively.

**Table 1. A summary of cumulative hospitalization rate for each age group**

| Age Group | Hospitalization rate per million |
|---|---|
| Overall | 94.5 |
| 0-4 years | 7.4 |
| 5-17 years | 3.5 |
| 18-49 years | 56.5 |
| 50-64 years | 143.0 |
| 65+ years | 286.9 |

The upsurge of positive cases poses significant challenges on the hospital capacity. As shown in Table 2, as of June 18, 2020, 70% of inpatient beds are occupied, in which 5% is used for COVID-19 patients. Also, nearly 63% of intensive care units (ICU) beds are occupied [1]. In addition, the shortages of medical supplies (e.g., personal protection equipment (PPE)) become more and more prevalent in the health systems with a rising number of coronavirus cases and hospitalizations. In the era of globalization, US medical supplies are heavily dependent on importation, nearly 72% of active pharmaceutical ingredients (APIs) are imported from other countries. Specifically, approximately 13% of medical products are from China, and 18% of pharmaceutical imports are provided by India [15]. Also, generic drugs imported from these two countries account for about 90% of medicine supplies in the US. However, the COVID-19 outbreak in January shuts down almost all manufacturing facilities and non-essential businesses in China. Even though manufacturing activities were resumed in late February, the average capacity utilization at top 500 manufacturing enterprises in China was only 58.98% [16, 17]. As such, a disrupted supply chain causes serious shortages of medical products in the US, which endangers the healthcare workers in the front line.

**Table 2. National estimates of hospital bed occupancy in the COVID-19 in the US**

| Estimates for June 18 | Number (95% CI) | Percentage (95% CI) |
|---|---|---|
| Inpatient Beds Occupied (all Patients) | 524,610 (500,844 - 548,376) | 65% (64% - 66%) |
| Inpatient Beds Occupied (COVID-19) | 40,112 (37,682 - 42,541) | 5% (5% - 5%) |
| ICU Beds Occupied (all Patients) | 77,029 (72,135 - 81,922) | 63% (61% - 64%) |

Indeed, healthcare workers are among the most vulnerable group of people who face a higher probability to get infected during the epidemic outbreak. The higher risk is due to their closer contact with patients, the shortage of PPEs, the delay of testing program in the early stage, and the high infection rate in the hospital. As the COVID-19 proliferates, healthcare workers suffer from occupational burnout and fatigue. The key factors include occupational hazards, emergence responses, process inefficiencies, and financial instability [18-20]. During the period of February 12-April 9, 2020, approximately 19% of COVID-19 patients are healthcare workers. Therefore, this fact further exacerbates the shortage of staffing in the hospital. To avoid secondary infection in the hospital, screening and masks are required for all people upon entry into the hospital [21]. Patients with suspected or confirmed COVID-19 are placed in a single-occupancy room with a closed door and a separated bathroom. Also, all healthcare workers should wear PPE, isolation gowns and non-sterile gloves upon entering these patients' room. When transporting patients out of the room, both patients and healthcare workers should wear PPE. Moreover, hospitals conduct routine cleaning and disinfection procedures. Enhanced environmental cleaning and disinfection are preferred for rooms used by patients with suspected or confirmed COVID-19, and for areas used by healthcare workers who care for such patients [22].

## B. Economic Challenges

The COVID-19 epidemic made the nation shut down non-essential businesses, schools and instituted travel bans, which have greatly impacted the U.S. economy. The shocks to supply chain bring significant disruptions to manufacturing. Small and medium manufacturing enterprises faced unprecedented challenges, while some have to shut down entirely to mitigate the virus spread. With social distancing measures in place, many workers can only work from home. The production lead time has doubled due to shortages of workers and materials [17]. Also, a limited number of products can be distributed worldwide by air or ocean because of trade wars, hiking tariffs, and importation restrictions. All these impacts of COVID-19 make companies question the just-in-time strategy and reconsider the design of supply chain. In March 2020, there was a 6.3% drop in manufacturing production, which was the largest one-month drop since 1946 [17, 23]. The drop was even larger for April 2020. Note that the Purchasing Manager's Index (PMI) shows the impacts of COVID-19 on the economy. PMI is a composite index, ranging from 0 to 100, of economic activities including new orders, inventory levels, production, supplier deliveries, and employment. If the PMI is above 50, the manufacturing sector is generally expanding. If PMI is below 50, it is generally contracting. As shown in Figure 2, US economic growth is strong in January 2020 with PMI 50.9, but decreases from January to April 2020 [17, 23]. When the COVID-19 outbreak occurred in March 2020, the PMI fell below 50, further dropped to 41.7 in April 2020, and then remained low through May 2020. From March to May 2020, COVID-19 poses significant challenges on the US economic activities due to unexpected outbreaks, lockdowns, and non-pharmaceutical interventions. After June 2020, the US economical activities recover with the rollout of stimulus plans, increasing manufacturing productions, and new modes for businesses such as teleconferencing, e-commerce and online learning.

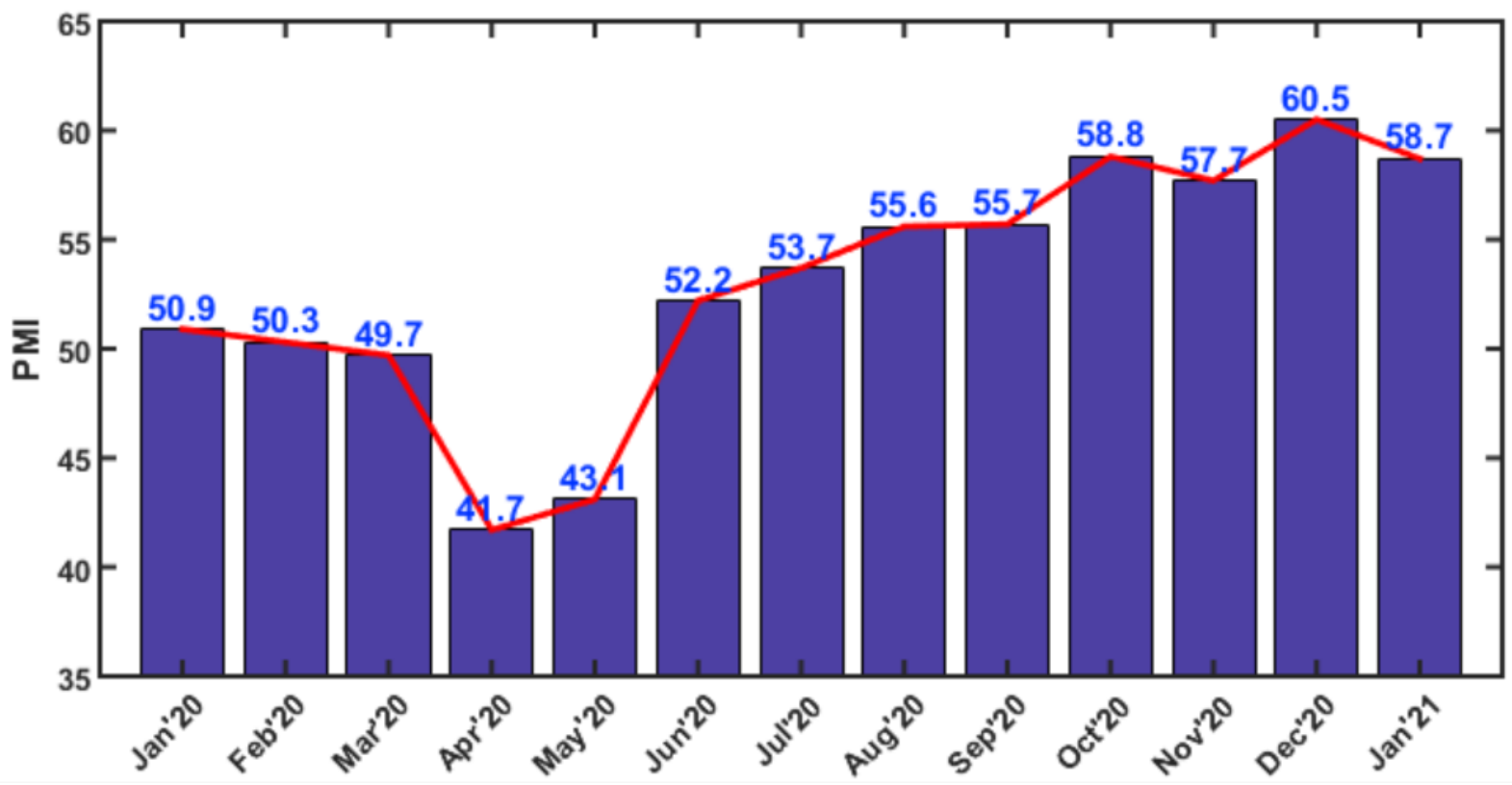

**Figure 2. The variations of Purchasing Manager's Index (PMI) from January to May 2020**

A worse impact on the manufacturing industry during the epidemic would be caused by decreased spending because of job loss or reduced incomes. The disruption in the manufacturing industry and the tremendous drop in demand led to the layoff of workers. As of May 2020, the unemployment rate in the manufacturing industry increased to 11.6%. Table 3 summarizes the number of employees in the manufacturing sector as issued by the U.S. Bureau of Labor Statistics, for both the non-seasonally adjusted case and the seasonally adjusted case [24]. As shown in Table 3, when it is not seasonally adjusted, the number of employees in the manufacturing sector decreased by 1.32 million from March 2020 to April 2020, with about 0.90 million in durable goods manufacturing and 0.42 million in non-durable goods manufacturing. Meanwhile, there were about 1.13 million fewer jobs in May 2020, compared to May 2019. When it is seasonally adjusted, the U.S. manufacturing lost about 1.29 million jobs from March 2020 to April 2020. About 69% (0.91 million) of the job loss was in the durable good manufacturing, while the rest 31% (0.38 million) was in the non-durable good manufacturing. Compared to May 2019, there were 1.12 million fewer jobs in May 2020 [24].

**Table 3. Employees on nonfarm payrolls in manufacturing (in thousands)** [24]

| | Not seasonally adjusted | | | | Seasonally adjusted | | | | |
|---|---|---|---|---|---|---|---|---|---|
| | May 2019 | Mar. 2020 | Apr. 2020 | May 2020 | May 2019 | Mar. 2020 | Apr. 2020 | May 2020 | Change Apr.– May 2020 |
| Manufacturing | 12,810 | 12,747 | 11,427 | 11,677 | 12,829 | 12,806 | 11,482 | 11,707 | 225 |
| Durable goods | 8,052 | 8,013 | 7,109 | 7,234 | 8,056 | 8,031 | 7,124 | 7,243 | 119 |
| Nondurable goods | 4,758 | 4,734 | 4,318 | 4,443 | 4,773 | 4,775 | 4,358 | 4,464 | 106 |

Schools and universities across the country have also been disrupted. In March 2020, most schools started to switch from in-person instruction to online-only instruction, which gave rise to the concerns about instruction quality [25]. Meanwhile, it is not uncommon that many universities faced financial challenges.

As students moved out of on-campus housing, universities issued pro-rated refunds to them, which was a substantial amount of unexpected expenses. Also, universities needed to allocate additional funds for dorm cleaning and technology essentials for online classes. Moreover, due to the cancellation of college entrance exams worldwide and limitation on travel, the enrollment for the fall 2021 semester is likely to drop, which will also cause financial issues to universities.

These paramount challenges posed by epidemics call upon multiple scientific disciplines to design and develop new enabling methods and technological innovations for rapid response and management. For example, a complete picture of the new virus is urgently needed from the community of medical scientists. The manufacturing community should be agile to innovate the design and increase the production of personal protective equipment (PPE). In this paper, we propose a system informatics approach for data-driven epidemic response and operational management, thereby mitigating the risks and controlling the virus spread. In the following sections, "**Measure**" provides statistical methods for optimal sampling and testing of the population for the presence of virus, as well as a review of data management and data visualization methods. "**Analyze**" focuses on the handling and analysis of heterogeneous and interconnected datasets (e.g., from CDC, Census Bureau, Food and Drug Administration, state and federal health departments) that are collected during the epidemic lifecycle. "**Improve**" exploits data-driven knowledge to improve the resilience design of health systems, including healthcare capacity, resources, workflows, and operations. Further, "**Control**" focuses on the learning and optimization of health policies and action strategies for controlling the spread of virus. The system informatics methods and tools will complement medical, clinical and pharmaceutical research efforts, helping safeguard the population from infectious diseases and make health systems more resilient to overwhelming epidemic events.

## III. Measure the Epidemic Dynamics

The "measure" step is directly aimed at testing the population for the prevalence of virus, which is critical to monitoring the temporal evolution of an epidemic in a spatial region. Rapid advances of gene, microbiology and imaging technologies have greatly improved the design and development of testing methods (e.g., speed and accuracy) of coronavirus and influenza. As discussed in Section II, an epidemic poses paramount challenges on the health and economy of our society. The prevalence of a virus in a large population often incurs large amounts of testing, which leads to spatially-temporally big data. This provides an opportunity for the "analyze" step to develop an in-depth understanding of dynamically evolving statuses of an epidemic. Here, data could be collected in disparate efforts by private companies, research centers, universities, and government agencies, thereby leading to the formation of data cohorts to address issues of data management. Epidemic data can then be visualized in various ways to provide comprehensible information about the spatiotemporal variations of an epidemic. An effective visualization further helps the "analyze" step to estimate and extract salient features for the prediction of future trajectory or the monitoring of transmission risks.

### A. Testing and Sampling

Clinical testing is a critical first step to stopping the spread, which consists of viral testing (i.e., examine whether an individual is currently infected or not) [26] and antibody testing (i.e., check whether an individual was infected before and currently has the presence of antibodies in the blood) [27]. In the case of COVID-19, specimens are often collected through swabs in the nose or throat for the viral testing. If specimens show the existence of a virus's ribonucleic acid (RNA) or proteins, the test will be positive. The antibody testing is typically done by collecting a sample of blood serum and then examining the presence of antibodies. In order to monitor the prevalence of virus, testing can be performed in three different ways as follows:

- **100% testing**: Population is the entire collection of individuals of interests in a region of interest (e.g., university, city, county, or state). If the cost is not a concern, 100% testing makes sure everyone is tested and then all the infected individuals can be isolated and quarantined. This is an effective approach to stop the spread, but often encounters practical limitations such as inadequate supply of testing kits, prohibitive cost, and population instability due to mobility and immigration.

- **Acceptance sampling**: Sample is a representative subset of the population that can be tested for statistical inference. Acceptance sampling, also called Lot Quality Assurance Sampling (LQAS) [28], is a middle ground between 0% and 100% testing and requires a small sample size for population surveys. The population can be stratified into sub-groups (or lots), and each lot can be sampled for clinical testing so as to “accept” or “reject” the lot according to the risk tolerance levels. Also, these samples can be aggregated to establish the confidence interval of infected proportion for testing the hypothesis on the prevalence of an epidemic virus.
- **0% testing**: This means that no testing will be done for the individuals in a specific region. In the onset of an epidemic, few tests are performed because the new virus is just emerging and has not caught enough attention from the public. Once the epidemic virus is captured (e.g., genome sequenced and shared), testing kits can then be designed and developed.

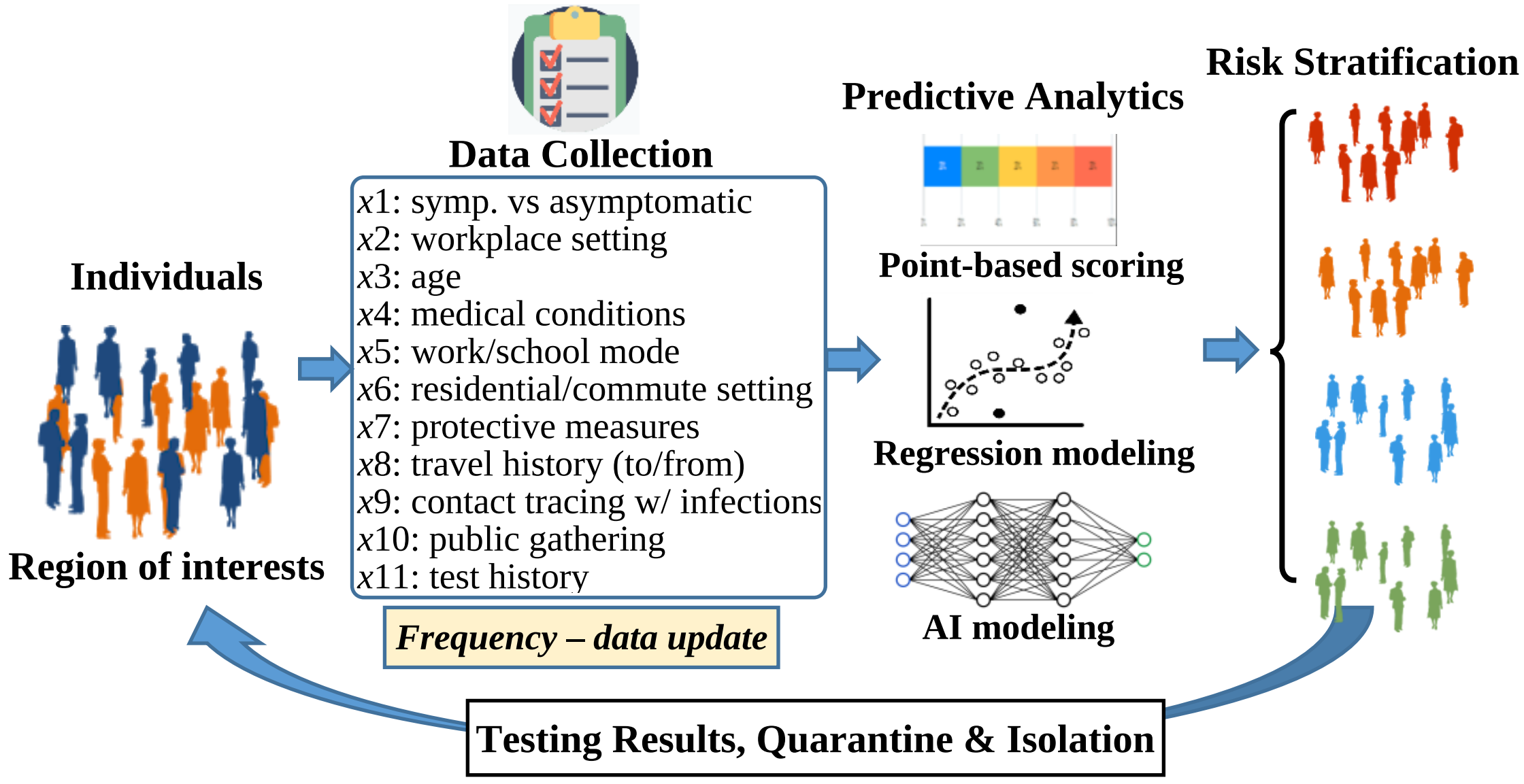

**Figure 3. Data-driven risk scoring systems for categorized sampling and testing**

Figure 3 shows that mobile or web-based applications can be used for data collection from individuals in a spatial region of interests, if the testing capacity is constrained and 100% testing cannot be implemented. Examples of the predictors may include $x_1$: symp. vs asymptomatic; $x_2$: workplace setting; $x3$: age; $x4$: medical/comorbidity conditions; $x5$: work/school mode; $x6$: residential/commute setting; $x7$: protective measures; $x8$: travel history (to/from); $x9$: contact tracing with infections; $x10$: public gathering; $x11$: test history; The response variable will be the risk probability of infection (range from 0 to 1). The data-driven decision support system helps stratify the individuals into groups (or lots) and then optimize the testing decisions. The risk scoring system categorizes the population into different groups with various levels of risk probability. For example, four groups can be stratified based on the risk probability, which helps further optimize the allocation of testing resources and identify the infected individuals for isolation and quarantine.

As shown in Figure 3, risk scoring systems can be established in three different ways, namely point-based systems, regression modeling, or AI-based modeling. Such scoring systems help categorize the acuity levels of patients and then improve the quality of healthcare services (e.g., surgical procedures, medication usages, care guidelines, treatment plans, and resource allocations) [29, 30]. Point-based scoring systems use the simple points or weights, and can be easily implemented in questionnaire form. The points or weights can be adjusted for different predictors (or factors). For example, if the symptom is weighted more than other predictors, it may be assigned with a larger point (or weight). In clinical practice, point-based scoring systems are widely used to stratify the patients, e.g., Acute Physiology and Chronic Health Evaluation (APACHE) [31], Sequential Organ Failure Assessment (SOFA) [32], Simplified Acute Physiology Score (SAPS) [33, 34], and Mini-mental state examination (MMSE) [35]. Figure 3 shows an example of risk factors for the design of point-based scoring systems, which also helps reduce the number of variables to compile into a short survey. An increasing score indicates a higher risk of infection. In addition, the infection risk can be derived using a multivariate logistic regression model as: $\log\left(\frac{risk}{1-risk}\right) = a + \sum_i b_i x_i$, where $Risk$ is the risk of death, $\left(\frac{risk}{1-risk}\right)$ is the odds ratio, $a$ is the intercept, $b_i$ is the coefficients and $x_i's$ are independent predictors). Here, training data or medical domain knowledge can be used to adjust the regression coefficients for different predictors (or factors). Finally, it is not uncommon that AI modeling (e.g., neural networks) are utilized to learn from complex-structured data for risk stratification. AI models, however, need large amounts of data for training and learning the weights, and are difficult to implement for testing and sampling in an epidemic.

Statistical sampling is a cost-effective approach to survey the groups (or lots) of individuals when the testing capacity and supply chain are constrained. First, the confidence interval for the proportion of infections $p$ can be estimated from testing data. If there are $c$ infected individuals for a random sample of size $n$, then an approximate $100(1-\alpha)\%$ confidence interval for $p$ is

$$\hat{p} - z_{\alpha/2}\sqrt{\frac{\hat{p}(1-\hat{p})}{n}} < p < \hat{p} + z_{\alpha/2}\sqrt{\frac{\hat{p}(1-\hat{p})}{n}} \quad (1)$$

where $\hat{p}$ is $c/n$, and $z_{\alpha/2}$ is the z value with an upper tail area of $\alpha/2$. This estimation tends to be more reliable when the number of confirmed individuals $c$ is greater than 6 in the sample, and is also applicable in the case of hypergeometric distribution when the sample size $n$ is small. Here, the choice of sample size is dependent on the significant level $\alpha$ and the margin of error (MOE), i.e., $z_{\alpha/2}\sqrt{\hat{p}(1-\hat{p})/n}$. If a specific MOE value $e$ is desired, then the sample size $n$ is approximately $z_{\alpha/2}^2\hat{p}(1-\hat{p})/e^2$. Note that the function $\hat{p}(1-\hat{p})$ reaches the maximum $1/4$ when $\hat{p} = 1/2$. Hence, the MOE is guaranteed not to exceed $e$ if the sample size is chosen to be $z_{\alpha/2}^2/4e^2$. For example, it is 95% confident that the MOE will not exceed 0.02 when the sample size is $1.96^2/(4 \times 0.02^2) = 2401$.

Acceptance sampling is useful to help the decision-making process on whether or not to lockdown or reopen a region (or "lot") for regular businesses. As shown in Figure 4a, the operating characteristic (OC) curve describes an acceptance sampling plan in terms of the probability of reopening versus the proportion infected. For example, the probability of reopening is $1-\alpha$ if the region meets the acceptance risk level (ARL) $p_{ARL}$. The probability of reopening is $\beta$ if the region is on the rejection risk level (RRL) $p_{RRL}$. Assuming a binomial distribution, the sample size $n$ and acceptable number $a$ can be obtained as:

$$1-\alpha = \sum_{c=0}^{a}\frac{n!}{c!\,(n-c)!}p_{ARL}^{c}(1-p_{ARL})^{n-c} \quad (2)$$

$$\beta = \sum_{c=0}^{a}\frac{n!}{c!\,(n-c)!}p_{RRL}^{c}(1-p_{RRL})^{n-c} \quad (3)$$

Then, for this acceptance sampling plan, if there are more than $a$ infections in the random sample of size $n$ from the region, lockdown will be implemented. If there are less than or equal to $a$ infections, the risk is below the ARL level and the region can be reopened. For example, Figure 4b shows the acceptance sampling plans with $n = 2000$ and $a$ is ranging from 15 to 95. When the acceptance number $a$ increases, this does not significantly change the slope, but rather move the OC curves to the right. If the acceptance number a is small, the risk tolerance levels tend to be low. For larger values of a, both ARL and RRL levels are higher. If a region is above the RRL, NPIs such as lockdown and stay-at-home should be implemented. On the other hand, rectification testing programs can further screen individuals in the rejected region. Often, 100% testing can be performed to identify all the infected individuals, then isolate and quarantine them.

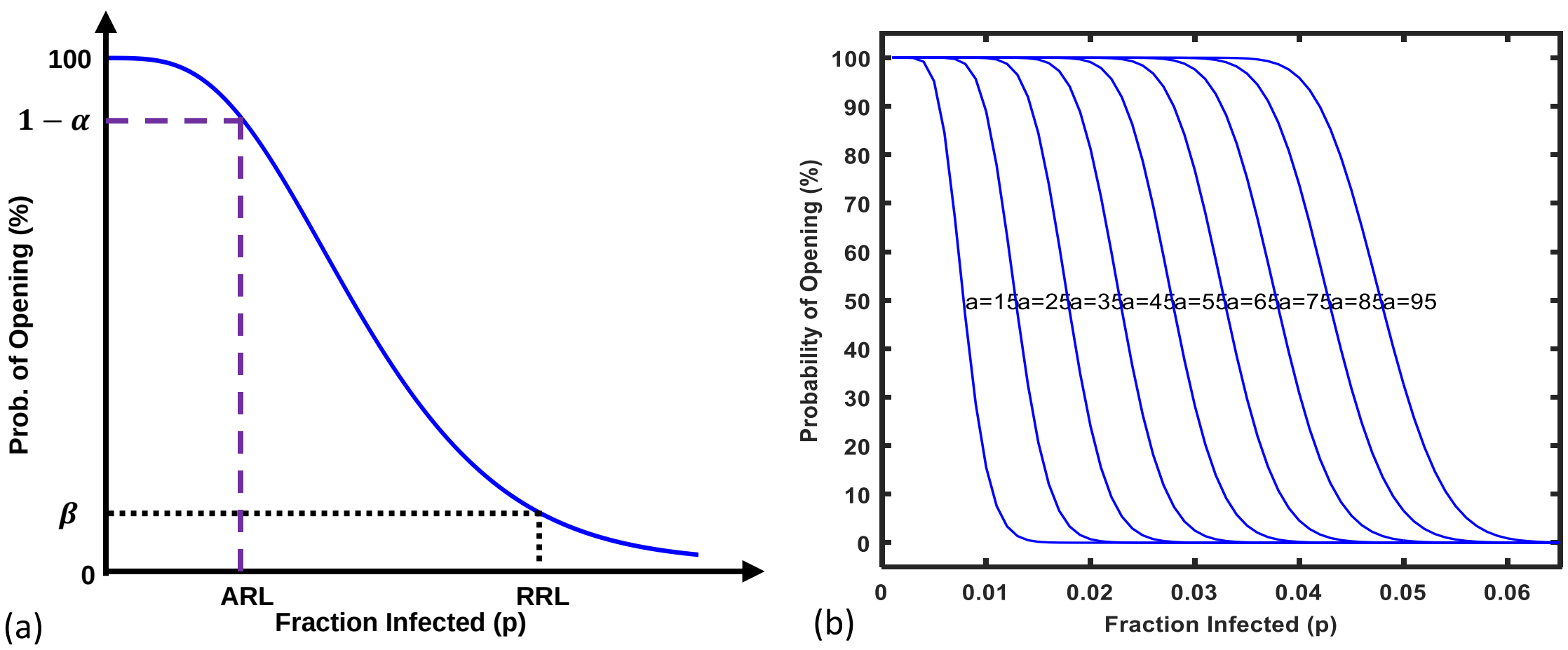

**Figure 4. (a) An illustration of operating characteristic (OC) curve, (b) OC curves of acceptance sampling plans with the sample size $n$=2000 and the acceptance number $a$ is ranging from 15 to 95.**

In the practice of clinical testing, acceptance sampling may have the following limitations. First, if the sample size is finite, then the distribution tends to be hypergeometric instead of binomial. However, binomial approximation of hypergeometric is valid if the ratio between sample size and lot size is less than 1/10. Second, acceptance sampling assumes the selection of samples at random from each region. Although clinical testing is prioritized for symptomatic cases or traced contacts of infected individuals, it can however assume that the infection of an individual is at random. Then, clinical testing can be assumed to be implemented on individuals who are infected at random, albeit with the introduction of bias to some extent. Third, individuals are assumed to be homogeneous in a region. In other words, homogeneity refers to the fact that the probability to get infected is approximately the same if in contact with pathogens. This is a reasonable assumption for a susceptible population, although there may be slight differences in the infection probabilities for uncontrollable factors such as age groups and blood types. These limitations and assumptions should be considered during the practice of acceptance sampling for clinical testing.

## B. Spatiotemporal Surveillance of Epidemic Processes

Clinical testing brings significant amount of data pertinent to the evolution of an epidemic. The epidemic data may include total cumulative cases (or per capita), daily new cases, total deaths for multiple spatial regions (or lots) of interest and are dynamically changing over time. Therefore, the epidemic evolution is a spatiotemporal process, i.e., varying in both space and time. The availability of data provides a great opportunity to design monitoring charts and develop epidemiology surveillance programs. Statistical monitoring methods help health systems leverage sequentially observed data to trigger the alarms and identify the outbreak region. However, raw data are often not normalized and cannot be directly used to develop monitoring charts. For example, spatial regions often have different population sizes. Total cases should be adjusted for the population in a region. As such, features need to be extracted from the data to describe the epidemic characteristics in a region. Examples of features may include cases per million, the incidence rate, or transmission risk index that are characterized with data-driven models.

If the monitoring objective is to detect abnormal changes of incidence rates $x_1, x_2, \ldots, x_k$ over $k$ regions, then the feature vector will be $\boldsymbol{x} = [x_1, x_2, \ldots, x_k]^T$. The statistical test is aimed at setting up the null and alternative hypotheses, then seeking data-driven evidence to determine whether an anomaly is present in any dimension (i.e., a region) of the feature vector or not. Under the null hypothesis $H_0$, the incidence rates over $k$ regions do not change over time. As such, the feature vector $\boldsymbol{x}$ is assumed to follow a multivariate normal distribution with population mean $\boldsymbol{\mu}$ and covariance matrix $\boldsymbol{\Sigma}$, i.e.,

$$f(\boldsymbol{x}) = \frac{1}{(2\pi)^{p/2}|\boldsymbol{\Sigma}|^{1/2}} \exp\left[-\frac{1}{2}(\boldsymbol{x}-\boldsymbol{\mu})'\boldsymbol{\Sigma}^{-1}(\boldsymbol{x}-\boldsymbol{\mu})\right] \tag{4}$$

If an outbreak occurs in one region or multiple adjacent regions, then the assumption of multivariate normal distribution is no longer valid. The alternative hypothesis $H_1$ that the joint distribution of multivariate features is non-normal will tend to hold. The hypothesis test accepts or rejects the null hypothesis $H_0$ at a significance level $\alpha$. Although the assumption of multivariate normality is required to formally establish confidence limits in the statistical test, a slight deviation will not severely impact the results [36]. Here, multivariate normal probability plotting can be used to evaluate whether the extracted features of incidence rates are approximately normally distributed for multiple regions of interests.

As shown in Figure 5(a), most of traditional monitoring schemes assume that $k$ regions are independent. Therefore, a common approach is to monitor each feature independently in the literature. In the bivariate case, control limits will form a rectangular region. If the pair of observations fall within this rectangular region, then the null hypothesis $H_0$ holds. If the pair of observations reside outside this region, then the null hypothesis $H_0$ is rejected. However, this monitoring scheme has limited applications due to the "curse of dimensionality". For example, if the probability of Type I error is α for each feature, then Type I error for monitoring $k$ features independently is $1-(1-\alpha)^k$. The probability that all $k$ observations fall within the confidence limits is $(1-\alpha)^k$ if all the $k$ regions are in control [37, 38]. Hence, the error is significant when the dimensionality of the feature vector increases. It may also be noted that $k$ features are oftentimes not independent because adjacent regions tend to be correlated with each other in an epidemic situation.

Therefore, multivariate statistical methods that consider spatial correlations and jointly monitor these regions (or features) are urgently needed. As shown in Figure 5(b), due to the correlation among adjacent regions, the pair of observations now resides in the elliptical region for the bivariate case. Under the null hypothesis $H_0$, $k$ regions will follow the multivariate normal distribution with the population covariance matrix $\boldsymbol{\Sigma}$. As such, the test statistic $\chi^2 = (\boldsymbol{x}-\boldsymbol{\mu})'\boldsymbol{\Sigma}^{-1}(\boldsymbol{x}-\boldsymbol{\mu})$ follows a chi-square distribution with $k$ degrees of freedom. The joint distribution changes in the presence of regional anomalies. If there are shifts in at least one out of $k$ regions, then $\chi^2$ values will be above the upper control limit $UCL = \chi^2_{\alpha,p}$, where $\alpha$ is the significance level. If $\chi^2$ values are below the upper control limit, then the null hypothesis $H_0$ holds and there will be no significant evidence of anomalies. The control ellipse of bivariate case in Figure 5(b) is due to region-to-region correlations. Because off-diagonal elements are no longer zero in covariance matrix $\boldsymbol{\Sigma}$, the principal axes of the ellipse are not parallel to the $\bar{x}_1$, $\bar{x}_2$ axes any more.

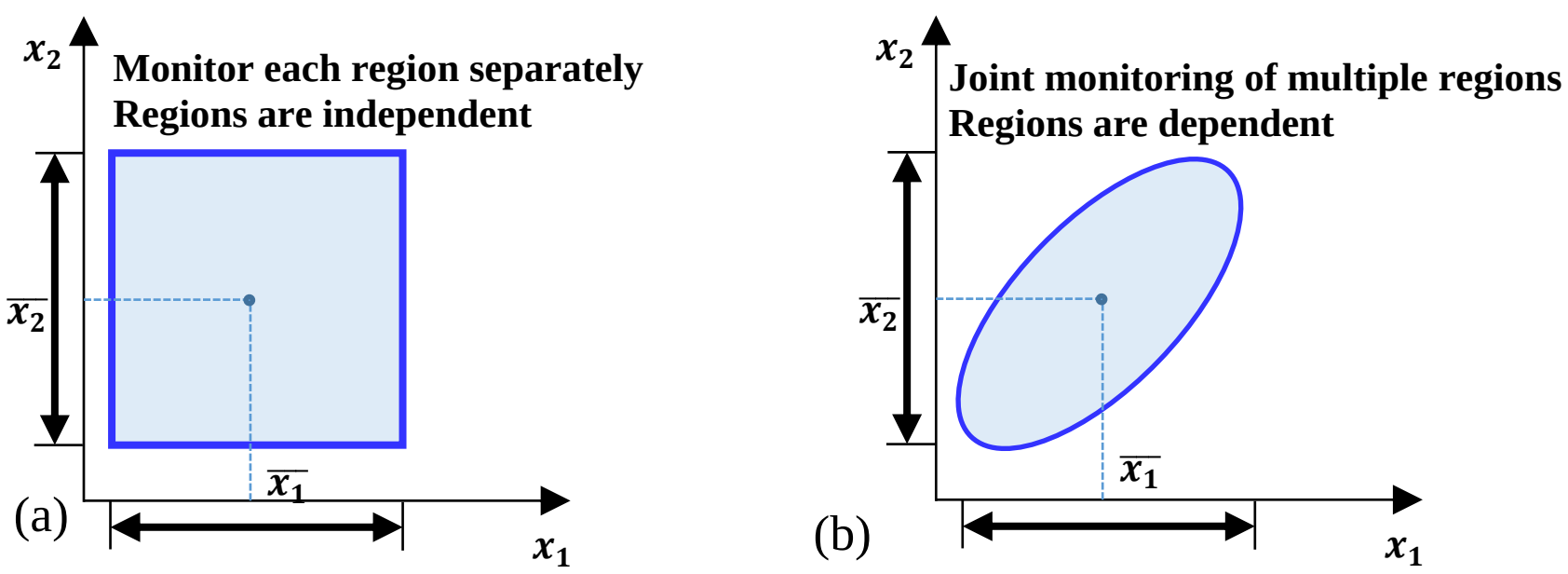

**Figure 5. Multivariate monitoring schemes for epidemic surveillance: (a) Monitor each region separately and regions are independent, (b) Joint monitoring of multiple regions and regions are dependent.**

In the real world, population mean $\boldsymbol{\mu}$ and covariance matrix $\boldsymbol{\Sigma}$ are often unknown and need to be estimated from the data. If the sample mean $\bar{\boldsymbol{x}}$ and covariance matrix $\boldsymbol{S}$ are used instead, then the test statistic becomes $T^2 = (\boldsymbol{x}-\bar{\boldsymbol{x}})'\boldsymbol{S}^{-1}(\boldsymbol{x}-\bar{\boldsymbol{x}})$, which is commonly called as the Hotelling $T^2$ statistic [39, 40]. The new UCL for the Hotelling $T^2$ statistic is:

$$UCL = \frac{p(N+1)(N-1)}{N^2 - Nk} F_{\alpha,k,N-k} \tag{5}$$

where $\boldsymbol{x}_1, \boldsymbol{x}_2, \cdots, \boldsymbol{x}_N$ are $N$ sequentially observed samples of epidemic data from $k$ regions, $F_{\alpha,k,N-k}$ is the upper $100\alpha\%$ critical point of $F$ distribution with $k$ and $N-k$ degrees of freedom. Note that control limits are established in Phase I with in-control datasets (i.e., without the presence of anomalies). For Phase II monitoring, the control chart plots control limits and the test statistic $T^2(i), i = 1,2, \cdots N$ for each sample. When a new sample arrives, we will then compute the test statistic and check the conformance in the control chart. Note that it is not feasible to graphically construct the control ellipse for more than two regions as shown in Figure 5b. The composite index (i.e., Hotelling $T^2$ statistic) helps characterize the multivariate distribution of $k$ features (or regions), and further establish the control chart to effectively detect whether there are shifts in at least one out of $k$ regions (i.e., multivariate epidemic monitoring and surveillance).

**Table 4. Examples of COVID-19 Data Repository/Cohort and Features**

| Data Cohorts and Repositories | Descriptions and Features |
|---|---|
| Center for Disease Control and Prevention (CDC) https://www.cdc.gov/coronavirus/2019-ncov/cases-updates/ | US infection data with cases, race, ethnicity, testing, hospital capacity and other data streams at local, state, and national levels |
| World Health Organization https://www.who.int/ | Global case updates with total confirmed cases and deaths, new cases and deaths, and transmission classifications |
| European CDC https://www.ecdc.europa.eu/en/covid-19-Epidemic | COVID-19 situation updates, case counts and distributions for the EU/EEA, UK, and worldwide. |
| National Institutes of Health https://datascience.nih.gov/covid-19-open-access-resources | COVID-19 data and resources such as official data, related studies, and high-performance computing consortium |
| National COVID Cohort Collaborative (N3C) https://cd2h.org/ | A very large patient-level COVID-19 clinical dataset shared by CTSA, CD2H and other distributed clinical data networks |
| Clinicaltrials.gov https://clinicaltrials.gov/ct2/results?cond=COVID-19 | Detailed information about active and recruiting clinical trials such as intervention and phase |
| Johns Hopkins University https://github.com/CSSEGISandData/COVID-19 | Global and US daily situation update at country and state level, along with time-series summary |
| NSF COVID Information Commons https://covid-info-commons.site.drupaldisttest.cc.columbia.edu/ | Open website to facilitate knowledge sharing and collaboration focused on NSF funded COVID rapid response research projects |
| New York Times https://github.com/nytimes/covid-19-data | US state level and county level situation updates, with historical and live data |
| Twitter Dataset https://github.com/thepanacealab/COVID-19_twitter | Tweets and retweets data acquired from Twitter stream related to COVID-19 chatter with all languages |
| The COVID Tracking Project https://covidtracking.com/data | US infection data with cases, tests, hospitalized, severity (in ICU, on ventilator, etc.), and outcomes |
| CORD-19 https://www.kaggle.com/allen-institute-for-ai/CORD-19-research-challenge | A dataset of over 167,000 scholarly articles about COVID-19, SARS-CoV-2 and related coronavirus |
| Ding Xiang Yuan https://ncov.dxy.cn/ | Global case updates with active, confirmed, recovered. China regional case updates with city level native/imported counts |
| OPENICPSR https://www.openicpsr.org/openicpsr/search/COVID-19/studies | Data cohort which contains links to US state policy database, government response dataset, and COVID-19 impact survey |

## C. Data Management and Visualization

As the epidemic progresses, large amounts of data are organized in the form of data cohorts or lakes. Medical scientists collect pertinent data about the clinical picture of a new virus for the development of effective intervention methods, such as antivirals and vaccines. Epidemiologists and engineers leverage the public health data to develop analytical models for the prediction of virus spread dynamics. Real-time data of epidemic situations is critical to understand the spread, trace the contacts, and control the propagation. Data management is indispensable to integrate disparate data efforts from government agencies, universities, and private companies. Here, data cohort connects various organizations to manage the data using the defining characteristics, which help researchers save tremendous amount of time in finding, analyzing, evaluating and validating relevant data for useful information and insights to stop the epidemic. Nonetheless, data lake is a repository of unorganized data in the raw format. Data cohort may include necessary data from on-going and completed research, as well as contact tracing data. This type of data could contain the patient location, sociodemographic information, and the list of contacts during the elicitation window and where the patient has visited. When the number of infections become prevalent, data management gets increasingly complex. This is partly due to the large number of cases, as well as the long list of traced contacts of each positive case. Data management depends on the use of database systems to support such many-to-many relational tables and provide a higher level of flexibility of routine data storage, update, security, reporting, and On-Line Analytical Processing (OLAP).

Note that the epidemic data is varying in both space and time. Table 4 provides examples of data repositories and cohorts developed by government agencies, institutions, and private companies. These data cohorts are open access to the public or limited access by applications. The UN data lab, US CDC and European Centers for Diseases Control (ECDC) organize and publish the real-time position data of virus spread in either country level or county level. Such information can be used to study and track the spread of the disease. US National Science Foundation (NSF) supported a research project to develop the COVID Information Commons, which is an open website to promote data and knowledge sharing across different COVID research efforts. National Institute of Health (NIH) initiated an National COVID Cohort Collaborative (N3C) project for collaboration on data collection, sharing, and analytics, which also provides the open access to research literature about COVID-19 genomics, virus structures, and clinical studies.

**Table 5. Examples of COVID-19 Visualization Dashboards and Features**

| Data Visualization Dashboard | Descriptions and Features |
|---|---|
| Center for Disease Control and Prevention https://www.cdc.gov/covid-data-tracker | US infection maps with testing outcomes, forecasting, demographic trends of sex, race/ethnicity and age, and social impacts |
| World Health Organization https://COVID-19.who.int/ | Country-level visualization of global trend of new cases, confirmed cases, and deaths |
| European CDC https://www.ecdc.europa.eu/en/covid-19-Epidemic | Interactive dashboard to show situation updates and case distributions for the EU/EEA, UK, worldwide. |
| Johns Hopkins University https://coronavirus.jhu.edu/us-map | Interactive visualization of confirmed cases, deaths, and a status health report in the US and across the globe |
| Institute for Health Metrics and Evaluation https://COVID-19.healthdata.org/ | Graphical visualization of deaths, infections and testing, and hospital resource utilization, predictions, and social distancing by country |
| Google https://news.google.com/COVID-19/map | A very high-level report of confirmed cases, recovered, deaths, and new cases (last 60 days) by country and worldwide |
| Facebook https://covid-survey.dataforgood.fb.com/ | Interactive visualization of infection proportion, population density, and elderly population by country |
| Bing https://www.bing.com/covid | Visualization of confirmed cases, recovered, deaths, and relevant news by counties in the US |
| Worldometer https://www.worldometers.info/coronavirus/ | Reported cases, deaths, and rankings by country or continent. |
| ArcGIS COVID-19 Hub https://coronavirus-disasterresponse.hub.arcgis.com/ | Esri storymaps and visualization tools to create time-lapse animation of the spread and help guide decisions around health, racial, and economic equity |
| 1 point 3 arc https://coronavirus.1point3acres.com/en | Interactive dashboard with a summary of the infected cases, deaths, recovered, and fatality rate. |
| Pharmaintelligence https://pharmaintelligence.informa.com/resources/key-topics/coronavirus | Drug discovery and clinical trial visualization across the globe |
| The weather company https://accelerator.weather.com/bi/ | High-level visualization of confirmed cases by day, by region (the last 14 days), deaths, rate of spread, rate of deaths, spread over time. |
| Coronavirus3d https://coronavirus3d.org/ | SARS-Cov-2 protein structure visualization |
| NextStrain https://nextstrain.org/ncov | Genomic epidemiology of novel coronavirus by region (Asia, Europe, North America, South America, etc), or by host, age, sex |

Also, academic institutions such as John Hopkins University (JHU) and the University of Washington provides the organized COVID-19 data and popular dashboards for data visualization. This, in turn, greatly facilitates the general public in visualizing the spread and trend of epidemic, thereby promoting situational awareness. In addition, there are data cohorts from private companies and foundations that provide targeted information about the disease. For example, the COVID-19 tracking project assembles the testing data, hospitalization rates, treatment outcomes, race and ethnicity data for researchers to investigate the outbreak scale, the mortality rate, and regional effects of the disease. COVID-19 Open Research Dataset (CORD19) provides an application programming interface (API) to retrieve the infection data, research feed, and COVID related texts. This API can help researchers query data in a fast manner. Surgo Foundation provides the community vulnerability index, social distance tracking, and nurse sentiment data to help develop analytical methods and tools for epidemic response.

Large amounts of data are readily available from different sources. The next step is to visualize and represent the data so that useful information and salient features can be easily comprehensible by the audience. Data visualization focuses on compact representations of trends and patterns in the data with graphical methods and tools such as time series charts, density graphs, and heat maps. The human brain can perceive information in graphics and images better than pale texts or data tables. An effective visualization helps condense a thousand words in one picture. There are a variety of visualization tools to represent the data in different ways, but it is important to choose the right tool to balance visual appearances and hidden information. The artisan spirit and craftsmanship help design better visualizations for the target users.

Table 5 provides examples of visualization dashboards available online for COVID-19, including the URL links and features. Most dashboards are developed with geographical maps and applications such as ArcGIS, as well as the COVID-19 data in the United States and worldwide. Figure 6 shows an illustration of the infection map in the county level of US from April 29 to September 23, 2020. The number of counties is close to 3,141. Instead of pale numbers in the table, such a visualization quickly provides a sense of the current status and the virus spread across US counties. Spatial regions are often labeled with a color map or with markers whose sizes are proportional to the number of infected cases. This informs people quickly about the regions of interests and the current spread of virus in the world. The temporal variations are shown as trends about how the number of cases rises with respect to time. The dashboard can also include data-derived features such as incidence rate, case-fatality ratio, testing rate, and hospitalization rate. Examples of popular dashboards include the CDC, JHU, Google, Bing, and 1point3arc dashboards. Notably, ArcGIS Storymaps provide a visualization tool to depict how the disease is spread from a regional epidemic to pandemic in a time-lapsed manner. Pharmaintelligence visualizes the progress of drug discovery and clinical trials worldwide, which highlight endeavors that medical scientists made to control the epidemic.

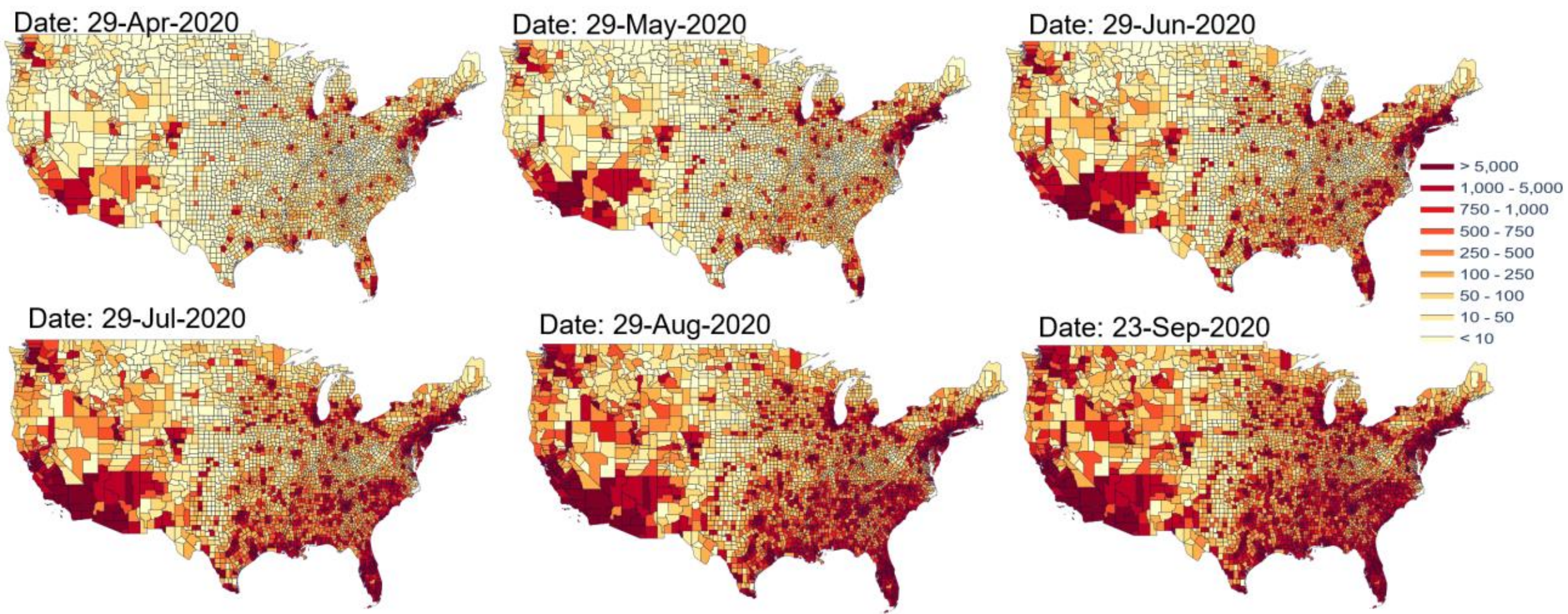

**Figure 6. The infection map of cumulative cases for 3,141 counties in the United States**

# IV. Analyze the Data for Epidemic Insights

The "analyze" step focuses on the extraction of useful information from epidemic data collected in the "measure" step. There are a variety of factors (e.g., demographics, socioeconomic factors, education factors, economy factors, population health factors, and mobility index) that may be interrelated with epidemic characteristics (e.g., the growth of confirmed cases). Therefore, it is critical to delineate and determine salient factors that are sensitive to the response variable. Note that the evolution of an epidemic is highly nonlinear and nonstationary. Traditional linear methods tend to be limited in their ability to handle the nonlinearity. High level of spatial heterogeneity also leads to skewed datasets and non-normal distributions of factors. As such, data transformation is necessary to pre-process and transform the data into normal shape. It is also imperative to utilize statistical models to investigate the interrelationships between various factors and epidemic characteristics. Also, rich data from the "measure" step can be fed into the development of simulation models. This, in turn, will help the "improve" step (see Section V) to forecast the real-time positions of virus spread and further run "what-if" analysis for the optimization of intervention strategies and healthcare policies. New experiments can then be designed to test the effectiveness of these action strategies on either physical systems or computer simulation models.

## A. Descriptive Analytics

This section of "descriptive analytics" aims to visualize the COVID-19 data and pertinent factors in an easily comprehensible form, and further investigate key predictors that are interrelated with the progress of infection situations in the US. In this study, the dependent variable (or responses) is set to be either cumulative $\boldsymbol{y_1}$ or weekly new cases $\boldsymbol{y_2}$ of COVID-19 infections at the county level[1], which are retrieved from New York Times data repository (i.e., https://github.com/nytimes/covid-19-data as shown in Table 4). To avoid confounding effects by population sizes, we have also considered response variables that are averaged by the population, i.e., cumulative $y_3$ or weekly new cases $y_4$ per capita in each county. The data repository provides real-time updates coronavirus cases in the US since January 2020, and provides cumulative daily counts of cases at state and county levels, respectively. We leveraged and processed the data at county level from Mar.29, 2020 (Week 1) to Aug.22, 2020 (Week 21) for the cumulative and incremental new cases of coronavirus at each week. In total, this study includes pertinent data about 2781 counties from 50 US states with a time span of 21 weeks (Mar.29-Aug.22, 2020) for the descriptive analytics.

For the independent variables, we have extracted a total of 72 predictors[2] at the county level from Google COVID-19 community mobility reports, US Census database, and County Health Rankings reports. Figure 7 depicts the geographical distribution of some example predictors (e.g., total population, household with grandparents living with grandchildren, people aged > 60) in 3,141 US counties. These predictors are categorized into four groups, namely social-economy, health, demography, and mobility, as follows.

- **Social-economic predictors:** We have extracted pertinent data about 3 education variables, 4 economic variables, and 9 occupational variables from the US census database in 2018 at the county level. (1) ***Education*:** The percentage of population aged 25 and over who don't have a degree ($x_1$), have a bachelor's degree or higher ($x_2$), and have a graduate or professional degree ($x_3$), respectively. (2) ***Economy:*** The unemployment rate among population aged 16 years and over ($x_4$), median household income ($x_5$), median family income ($x_6$) and median earnings ($x_7$). (3) ***Occupation:*** Among the employed population aged 16 years and over, we consider the percentage of population who work in management, business, science, and arts ($x_8$); service ($x_9$); sales and office occupations ($x_{10}$); natural resources, construction, and maintenance ($x_{11}$); production, transportation, and material moving ($x_{12}$); manufacturing ($x_{13}$); wholesale trade ($x_{14}$); retail trade ($x_{15}$); educational services, and health care and social assistance ($x_{16}$).
- **Health predictors:** Moreover, we extracted the data about 10 health features from County Health Rankings reports as follows: percentage of population with disability ($x_{17}$), percentage of adults that report fair or poor health ($x_{18}$), average number of reported physically unhealthy days ($x_{19}$), percentage of adults that reported currently smoking ($x_{20}$), food environment index ($x_{21}$), percentage of adults that report no leisure-time physical activity ($x_{22}$), percentage of the population with access to places for physical activity ($x_{23}$), percentage of adults that report excessive drinking ($x_{24}$), percentage of people under age 65 without insurance ($x_{25}$), and primary care physician (PCP) rate ($x_{26}$).

---

[1] Note that the distributions of response variables are highly skewed, and are therefore transformed to the log scale for descriptive analytics, i.e., $y' = \log(y + 1)$.
[2] If predictors are approximately normally distributed, no transform is made. For positively skewed data, $\log(x + 1)$ is used, while for negatively skewed data, $\log(\max(x + 1) - x)$ is applied.

- **Demography predictors:**

  ***Population:*** total population ($x_{27}$)
  ***Age and sex:*** median age ($x_{28}$), percentage of population aged >60 ($x_{29}$), and sex ratio (males per 100 females) ($x_{30}$).
  ***Household and family:*** average household size ($x_{31}$), average family size ($x_{32}$), number of households with grandparents living with grandchildren ($x_{33}$), and percentage of single-parent households ($x_{34}$).
  ***Marital status:*** percentage of population married ($x_{35}$), divorced ($x_{36}$), widowed ($x_{37}$).

- **Mobility predictors:** (1) ***Commute mode:*** The US Census provides the percentage of population aged 16 and over who drive alone ($x_{38}$), or carpool ($x_{39}$) by car, truck, van, or use public transportation (excluding taxicab) ($x_{40}$), as well as the percentage of workers who commute in their car alone commute more than 30 minutes ($x_{41}$), work at home ($x_{42}$). (2) ***Community mobility***: Google provides the community mobility change from the baseline (in percentage) in 6 different types of places, namely retail and recreation, groceries and pharmacies, parks, transit stations, workplaces, and residential. The mobility data provide insights about how COVID-19 and related policies impact the population's mobility patterns in public places. The data were organized on a daily basis. For each week from Mar.29 to Aug.22, we calculated the average, median, variance, skewness, kurtosis of mobility variations in each type of place at the county level, thereby extracting a total of 30 mobility features (i.e., $x_{43}{\sim}x_{72}$: 5 features × 6 places).

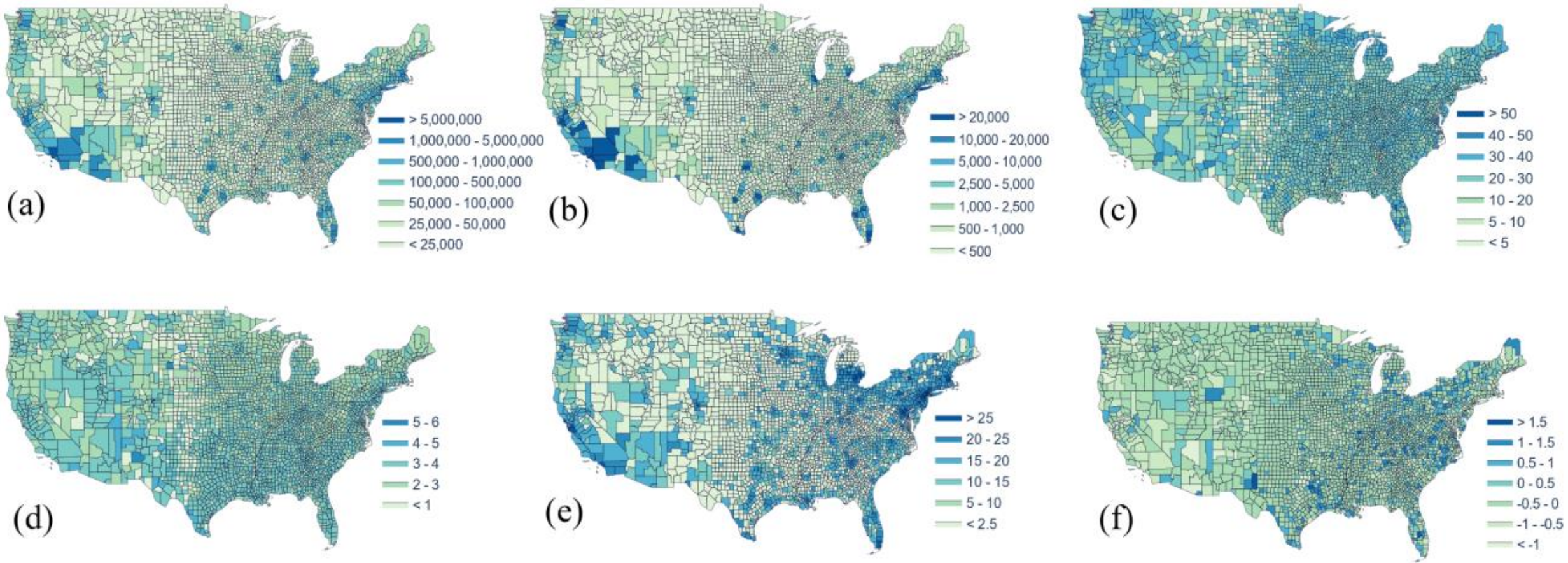

**Figure 7. The geographical distribution of (a) total population, (b) household with grandparents living with grandchildren, (c) percentage of people age > 60, (d) average family size, (e) mean of mobility change in residential setting (Apr.5) (f) skewness of mobility change in residential setting (Apr.5).**

### A.1 Correlation Analysis

Figure 8 (a) shows the Pearson correlations between 72 predictors (i.e., $x_1{\sim}x_{72}$, see details above) and cumulative confirmed cases $y_1$. In general, there are high correlations between COVID-19 situations and social-economic, demography and mobility predictors. The highest correlation (83.89%) is with the total population $x_{27}$ in each county, also see the scatter plot in Figure 9 (a). This shows the prevalence of COVID-19. The more population a county has, the more infections it will have. As of August 22, 2020, COVID-19 had spread over the whole US territory and few counties could be an exception.

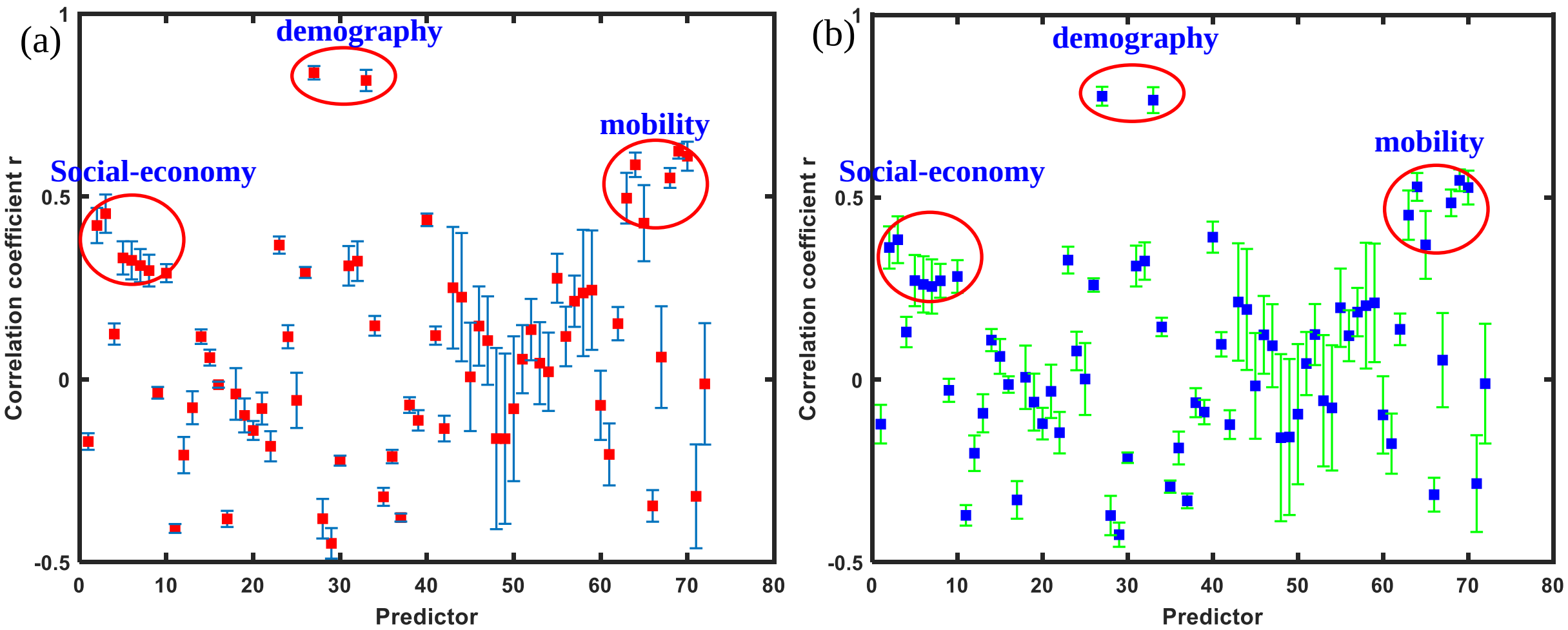

**Figure 8.The correlation between 72 predictors (i.e., $x_1 \sim x_{72}$) and cumulative confirmed cases $y_1$ (a) and weekly new cases $y_2$ of COVID-19.** The error bar represents the mean and standard deviation over 21 weeks.

The second highest (81.77%) is with the number of households with grandparents and grandchildren $x_{33}$; also see the scatter plot in Figure 9 (b). Elderly people and children are both high-risk groups. When the number of households with grandparents living with grandchildren is high, these two groups of people are more vulnerable and more likely to transmit the virus to each other.

For some social-economic factors, the Pearson correlations are approximately in the range of 29% to 46%, also see Figure 8. This is not as highly correlated as two demographic variables, but are sensitive to COVID-19 situations to some degree. Also, it may be noted that the Pearson correlations are approximately in the range of 42% to 63% for some mobility predictors. This is not surprising because the virus spread causes many businesses to shut down and people to stay at home. The variations of mobility patterns in community places are sensitive to the COVID-19 situations.

Further, we computed and compared with the Pearson correlations between 72 predictors (i.e., $x_1 \sim x_{72}$, see details above) and weekly new cases $y_2$ for 2781 counties from 50 US states, as shown in Figure 8 (b). The results are similar to the cumulative confirmed cases in Figure 8 (a), but with slight decreases in the magnitude of 3% to 10%. In other words, there are slightly higher correlations between 72 predictors and cumulative cases than weekly new cases.

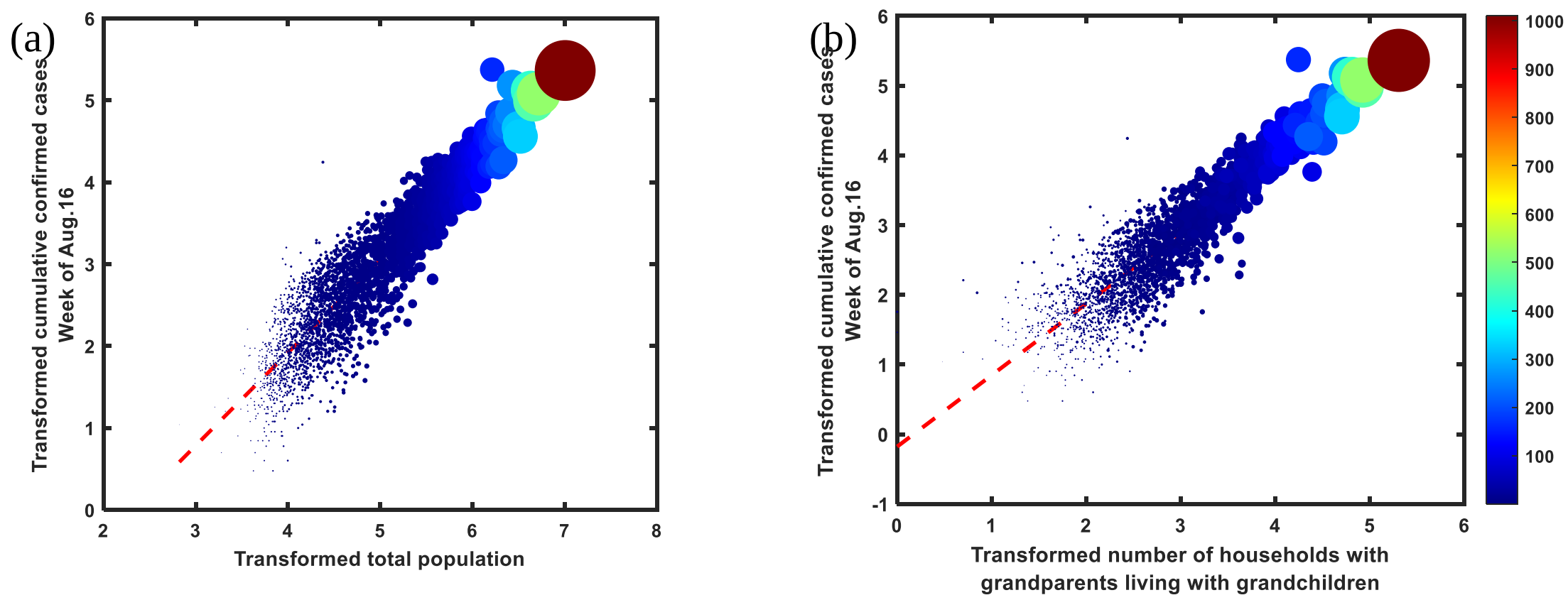

**Figure 9. Scatter plots of cumulative COVID-19 cases vs. total population (a) and the number of households with grandparents living with grandchildren (b).** The circle size is proportional to the county population.

The prevalence of coronavirus in the US leads to the highest correlation (i.e., 83.89%) with total population. This is conducive to building a regression model to forecast the growth of COVID-19 cases in each county. However, total population poses a confounding effect that dilutes the factorial effects from other predictors. Therefore, we have further examined each predictor's correlation with cumulative and weekly new cases per capita as response variables (i.e., $y_3$ and $y_4$, respectively), as shown in Figure 10.

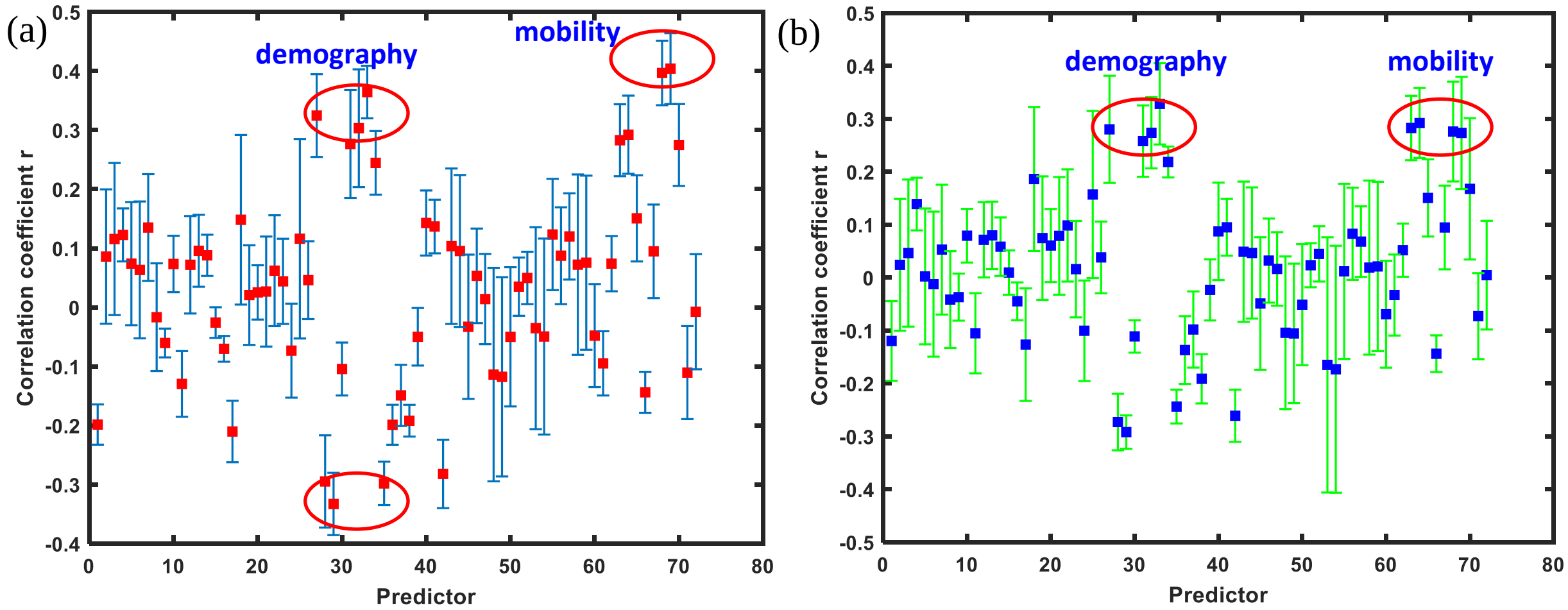

**Figure 10.The correlation between 72 predictors (i.e., $x_1 \sim x_{72}$) and cumulative confirmed cases per capita $y_3$ (a) and weekly new cases per capita $y_4$ of COVID-19.** The error bar represents the mean and standard deviation over 21 weeks.

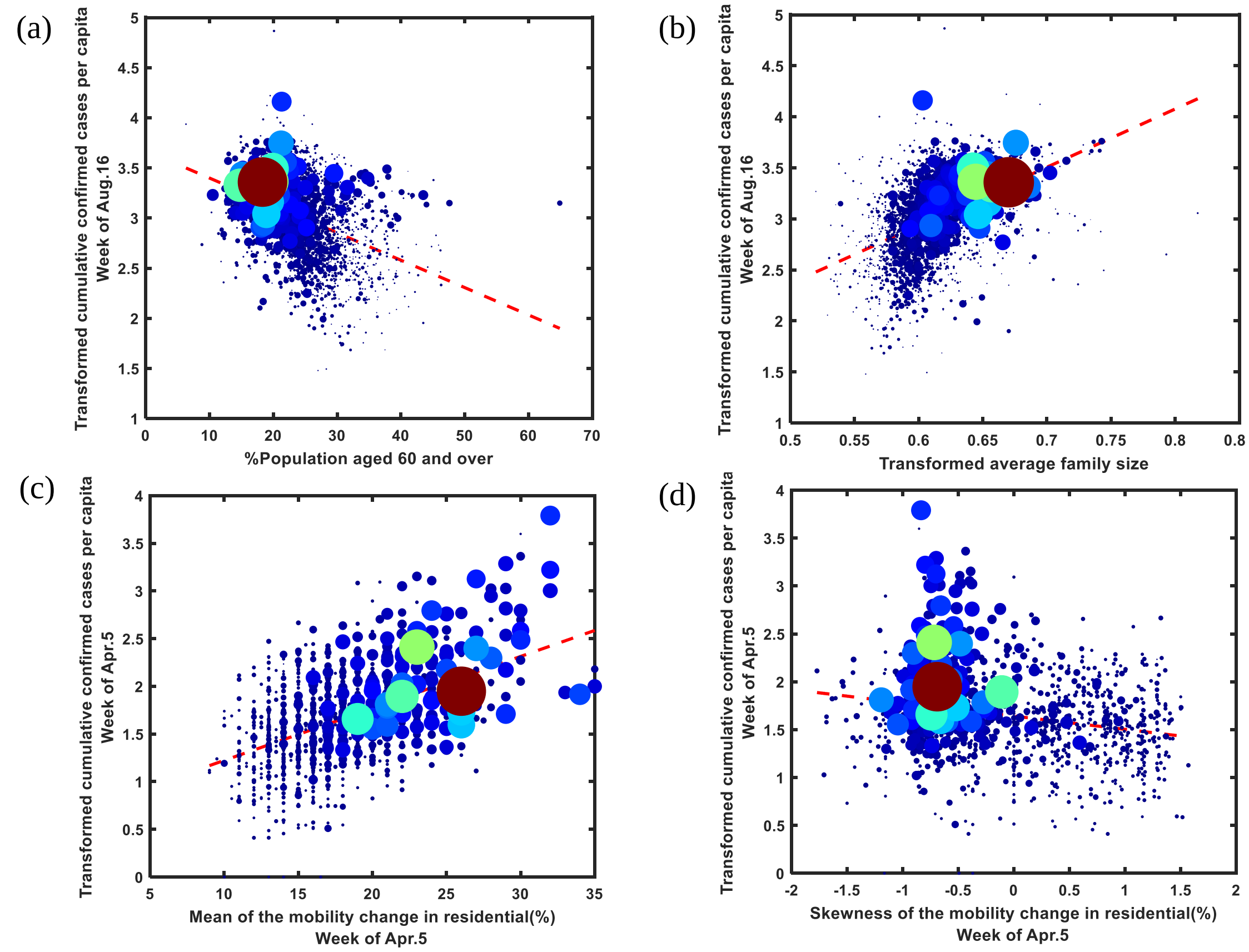

**Figure 11. Scatter plots of cumulative cases per capita vs. the percentage of the population aged >60 (a) the average family size in (b), the mean of mobility change in residential (%) (c), and the skewness of mobility change in residential (%) (d).** The circle size is proportional to the total population in each county.

Figure 10 shows that social-economic factors are no longer as significant as in Figure 8, and yield the average Pearson correlations below 20% after the per capita adjustment. Nonetheless, demographic and mobility predictors are still significant among all, although their Pearson correlations are approximately in the range of 30% to 40%. As shown in Figure 10 (a), the predictors with high correlations with cumulative confirmed cases per capita $y_3$ include: $x_{29}$ the percentage of the population aged >60 $x_{29}$ (-33.27%), the average family size $x_{32}$ (30.32%), the mean of mobility change in residential $x_{69}$ (40.4%), and the skewness of mobility change in residential $x_{68}$ (39.67%). The scatter plots in **Error! Reference source not found.** also show that there are correlations between these four predictors and the response variable (i.e., cumulative confirmed cases per capita $y_3$). However, neither positive nor negative correlations are as strong as the level of 83.89% in Figure 9. Similarly, Figure 10 (b) shows the Pearson correlations between 72 predictors and weekly new cases per capita $y_4$ for 2781 counties from 50 US states. The results are similar to the cumulative cases per capita in Figure 10 (a), but with slight decreases. In other words, weekly new cases per capita $y_4$ are essentially the week-by-week differences of $y_3$. Thus, there are slight decoupling of correlation effects.

### A.2 Regression Modeling

Section A.1 focuses on the relevancy between predictors and response variables. However, there is also redundancy (or multicollinearity) among the predictors that causes the regression model to be unstable and sensitive to external noises. A total of 72 predictors tend to bring the "curse of dimensionality" problem, and cause overfitting to the model. Therefore, we utilize the lasso regression model to shrink the number of predictors and further select a sparse set of significant variables. For a given value of $\lambda$, a nonnegative parameter, lasso regression penalizes the sum of L1 norm of regression parameters as:

$$\min_{\beta_0,\boldsymbol{\beta}} \left(\frac{1}{2N}\sum_{i=1}^{N}(y_i - \beta_0 - \boldsymbol{X}_i{}^T\boldsymbol{\beta})^2 + \lambda\sum_{j=1}^{p}|\beta_j|\right) \tag{6}$$

where $N$ is the number of observations, $y_i$ is the response at observation $i$, $\boldsymbol{X}_i$ is a vector of predictor values at observation $i$, and $p$ is the dimensionality of predictors. Lasso-penalized regression addresses the multicollinearity issue via regularized learning. A parsimonious set of predictors also helps increase the model interpretability, as opposed to a lower level of interpretability with the use of traditional dimensionality reduction methods (e.g., principal component analysis).

Figure 12 (a) shows the variations of prediction errors with respect to the regularization parameter $\lambda$. The lasso experiment is performed with 10-fold cross validation for the response variable of cumulative cases per capita and 72 predictors. When $\lambda$ decreases, the number of selected predictors increases. Note that the prediction error decreases to a local minimum and then increases. The optimal penalization parameter $\lambda_{\text{opt}}$ is identified at the location with minimal cross-validation error plus one standard deviation, which is as shown in Figure 12 (a) as the green dashed line and the green circle. For cumulative cases per capita, $\lambda_{\text{opt}}$ suggests the inclusion of 19 predictors which yield the lowest cross-validation error. It is evident that variable selection via Lasso penalization yields not only a sparser model, but also a smaller cross-validation error.

Figure 12 (b) shows the coefficient paths of 72 predictors when the value of $\lambda$ decreases. It may be noted that more and more predictors are included when $\lambda$ decreases. The green dashed line locates an optimal regularization parameter for the selection of 19 predictors that is identified using the 10-fold cross validation. We have repeated the experiments for each of four response variables ($y_1 \sim y_4$). The results are consistent with slight deviations because of the variations of correlations as in Figure 8 and Figure 10.

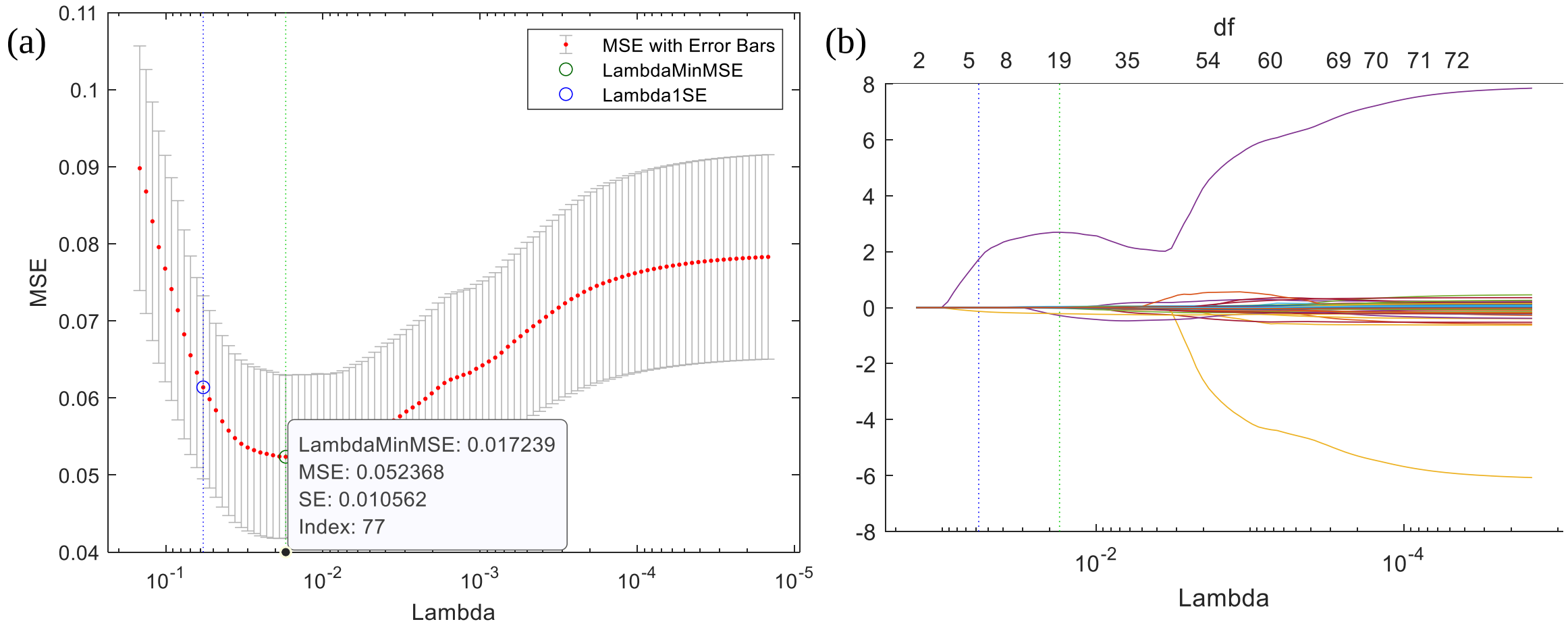

**Figure 12. (a) The variations of prediction errors vs. regularization parameter in Lasso regression with cumulative cases per capita and 72 predictors; (b) The coefficient path for the Lasso regression.**

Furthermore, we use the selected set of 19 predictors to build the fixed-effect regression models and investigate the relationship between predictors and the temporal variations of four response variables ($y_1 \sim y_4 | t$) over 21 weeks. The fixed-effect regression model is formulated as follows:

$$y_i | t = \beta_0 + \sum_{j=1}^{n} \beta_j x_{ij} + \sum_{k=1}^{m} \lambda_k I_{ik} + \varepsilon, \qquad t = 1,2,\cdots,21 \tag{7}$$

where $y_i$ is the number of cumulative (or weekly new) confirmed cases in county $i$, $x_{ij}$ is county $i$'s predictor $j$, $n$ is the total number of predictors, and $\beta_0$ and $\beta_j$ are parameter estimates. Also, $\lambda_k$ is the fixed effect for state $k$, $m$ is the number of states considered in the analysis, $I_{ik}$ is an indicator function for county $i$ and $I_{ik}$=1, if $i \in k$ (county $i$ belongs to state $k$); otherwise, $I_{ik}$=0. We fitted the regression model on a weekly basis.

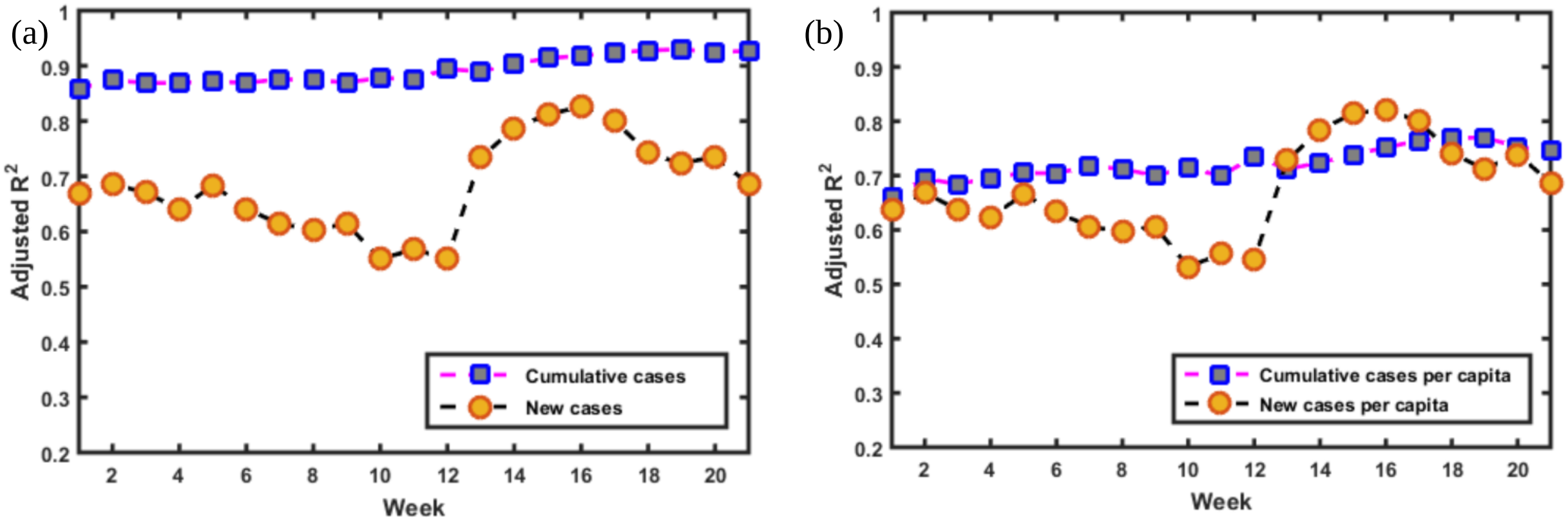

**Figure 13. The variations of adjusted $R^2$ for the fixed-effect models with response variables of (a) cumulative and weekly new cases, and (b) cumulative and weekly new cases per capita over the period of 21 weeks.**

Figure 13 (a) shows the variations of adjusted $R^2$ for the fixed-effect models with response variables of cumulative and weekly new cases, respectively, over 21 weeks. Note that the presence of highly sensitive predictors (e.g., total population, 83.89% correlation) achieves the adjusted $R^2$ in the range from 88% to 94% for cumulative cases. However, for weekly new cases, there is a high level of fluctuation in the adjusted $R^2$ (i.e., approximately 55% to 82%) over 21 weeks. This is mainly due to policy adjustments from local and federal governments (e.g., reopen the economy), causing high variations of weekly new cases. Also, these policies are not consistent and sometimes heterogeneous in different US counties. Nonetheless, the high adjusted $R^2$ values show the predictability of fixed-effect models. Figure 13 (b) shows the variations of adjusted $R^2$ for the fixed-effect models with response variables of cumulative and weekly new cases per capita, respectively, over 21 weeks. The results are consistent with correlation analysis in Section A.1. Because of the decrease in variable correlation, the adjusted $R^2$ values are approximately in the range from 71% to 79% for cumulative cases per capita. Similarly, for weekly new cases per capita, the adjusted $R^2$ are still fluctuating due to policy adjustments.

Figure 14 shows an example of residual plots that provide diagnosis results of the fixed-effect regression model with the response variable of cumulative cases per capita. Note that no systematic patterns are discerned in the residual plots. The histogram plot in Figure 14 (a) shows that the normality assumption is valid. Figure 14 (b) shows parallel bands centered around zero in the series of residuals.

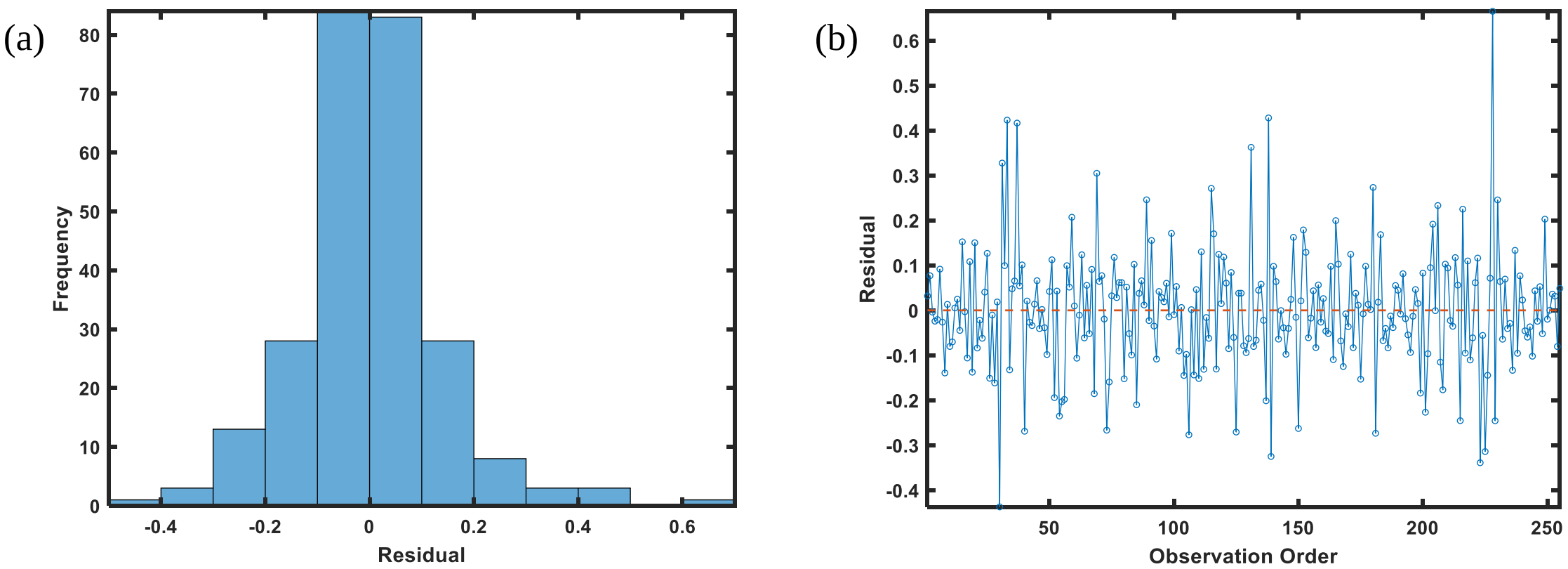

**Figure 14. The residual diagnosis of regression model with the response variable of cumulative cases per capita.**

## B. Spatiotemporal Analytics

The outbreak of an epidemic is often spatially distributed and evolves over time, thereby generating spatially and temporally big data. For example, epidemic situation reports, in days, months and even years with multiple waves of infections, brings about large amounts of data. Infection dynamics can be visualized through data dashboards or time-lapsed visualization in geographically distributed regions, e.g., see Section III.C. However, spatiotemporal data poses significant challenges for human experts to delineate key factor-to-factor interactions and predict the evolution of an epidemic. Fully utilizing the spatiotemporal data depends to a great extent on the development and implementation of information-processing methodologies. Only with effective analytical methods and tools, we can then enable and assist (i) the identification of key factors that are highly correlated with the epidemic growth, (ii) the development of spatiotemporal models for epidemic prediction and risk assessment, and (iii) the provision of decision-support tools for resource planning and intervention strategies towards smarter healthcare services.

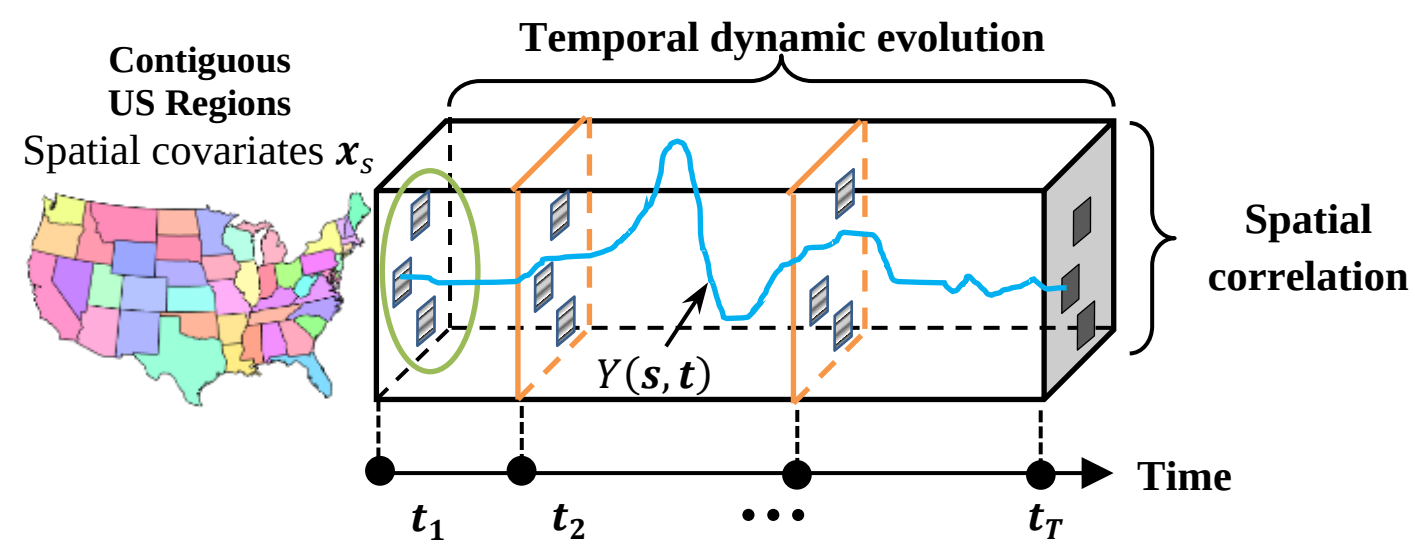

**Figure 15. An illustration of space-time dynamics in the evolution of an epidemic**

Figure 15 illustrates spatiotemporal dynamics of epidemic data generated over geographical regions in the contiguous US. Each cross-section is a snapshot of the epidemic situation at a particular time point. As the infection dynamics evolve across both space and time, epidemiological surveillance systems produce spatiotemporal data: $\{\mathrm{Y}(\boldsymbol{s}, t): \boldsymbol{s} \in \mathcal{S} \subset \mathbb{R}^d, t \in T\}$, where Y is dependent on both spatial domain $\mathcal{S}$ and time T symbolizes the spatiotemporal variations. Space and time dimensions are relevant but different in an epidemic. It may be noted that the time dimension includes the past, present, and future, which is not directly comparable to the space. Instead, the space dimension is indexed by spatial coordinates. Note that each spatial region can also be embodied with characteristic covariates, predictors or features, $\boldsymbol{x}_s$, such as demographics, socioeconomic factors, or mobility features. If two regions are close to each other, they tend to have a higher correlation. In general, the spatial "closeness" can be due to spatial distance, characteristic features, or high-level traffics (e.g., air transportation) between two regions. There is a need to investigate not only spatial correlation and temporal correlation, but also space-time interaction. Such spatiotemporal interactions bring substantial complexity in the scope of epidemic modeling and analytics.

In the past few decades, the proliferation of space-time data has fueled increased interests in spatiotemporal analytics. Examples of application areas include brain imaging [41, 42], public health [43, 44], service equity [45] and socio-economics [46]. The specific questions include the analysis of time-varying brain image and fMRI data, geographical diffusion of epidemic infectious diseases, and spatial equity of public services. Also, there are previous works that employ random fields in $\mathbb{R}^{\mathrm{d}+1}$ to model space and time dependencies [47]. After a review of the literature on spatiotemporal modeling, we summarize four classical models for the analysis of space-time indexed data below. These models are not meant to be comprehensive or exclusive, but rather serve as initial ideas for spatiotemporal modeling of an epidemic.

- **Spatially-varying time series model**: $Y(s,t) = Y_s(t)$, which separates the temporal analysis for each location. This model $Y_s(t)$ shows specific interests in time-dependent patterns for each spatial location, and allows for location-to-location analysis between time series. For example, the time series of infection cases can be represented and characterized at each zip code, county, or state to investigate the variations of health policy and pertinent impacts on each location. In the literature, Yang et al. extracted patterns from ECG time series at each sensor location on the body surface and exploited the useful information for the identification of cardiovascular diseases [48-50].

- **Temporally-varying spatial model**: $Y(s,t) = Y_t(s)$, which separates spatial analysis for each time point. The model $Y_t(s)$ focuses more on space-dependent patterns at a particular time point. For example, spatial patterns of virus spread can be modeled at a specific time point; then how spatial patterns change over time. Yao et al. studied body-surface ECG images during the period of ventricular contraction for the detection of myocardial infarction sites [51, 52]. However, both $Y_s(t)$ and $Y_t(s)$ are conditional methods that investigate either the space given time or time given space, which tend to be limited in their ability to capture space-time correlations.

- **Space-time separation model**: This model separates the spatial and temporal components in the multiplicative form as $Y(\boldsymbol{s},t) = M(\boldsymbol{x}_s)g(t)$, where $\boldsymbol{x}_s$ are the characteristic covariates for each spatial region, $Y(\boldsymbol{s},t)$ can be the number of cumulative confirmed cases for a spatial region $\boldsymbol{s}$ at time $t$. Here, $M(\boldsymbol{x}_s)$ can take the form of a nonlinear regression model form with adjusted fixed effects for each spatial region. The temporal growth $g(t)$ can be modeled with sigmoidal functions such as logistic or Gompertz functions. For example, Jia et al. presents a space-time separation model for the COVID-19 growth from Jan 24 to Feb 19, 2020 in China [53]. However, this separation model only accounts for multiplicative effects between spatial and temporal components, and can only model the exponential growth with saturation after a period of time.

- **Parameter-driven spatiotemporal model**: To increase the flexibility to model spatiotemporal dynamics, at a particular time point $t$, a spatial model can be developed for the cross-section data to represent how epidemic patterns are correlated with characteristic covariates $\boldsymbol{x}_s$, i.e., $Y(\boldsymbol{s},t) = M(\boldsymbol{x}_s;\boldsymbol{\beta}_t) + \varepsilon$, where $\varepsilon$ is the random noise and $M(\boldsymbol{x}_s;\boldsymbol{\beta}_t)$ is the parameterized model. As epidemic observations change over time, model parameters will also vary with respect to time, i.e., $\mathrm{M}(\boldsymbol{x}_s; \boldsymbol{\beta}_\mathbf{t}), \mathrm{M}(\boldsymbol{x}_s; \boldsymbol{\beta}_{\mathbf{t+1}}), \cdots$. Then, a state space model $\boldsymbol{\beta}_t = g(\boldsymbol{\beta}_{t-1}, \gamma)$ can be used to characterize temporal correlation and link the parameters over time, where $\mathrm{g}(\cdot)$ is the nonlinear evolution model and $\gamma$ is process noise. As such, spatial and temporal components interact with each other to sequentially update the model when new data are available at the next time point. For example, Yang et al. develop a sparse particle filtering approach for characterizing and modeling space-time dynamic data generated from stochastic sensor networks [54].

## C. Privacy-Preserving Data Analytics

As the epidemic data (e.g., contact tracing, quarantine) proliferate, people are increasingly concerned about privacy issues. When data resolution and dimensionality are high, each entry in a database is essentially unique. Hence, establishing a linkage with named individuals becomes a much simpler matter. In the traditional practice, data analytics tend to focus on the effectiveness and efficiency of models, but overlook privacy in the context of an epidemic. Privacy breaches can bring unexpected disruptions to health policies and mitigation efforts in the epidemic response. For example, data exfiltration of contact tracing endangers the privacy of pertinent individuals, thereby causing a trust crisis and potential failures to the execution of policies. It is estimated that healthcare systems suffer from the cost of approximately $300 billion annually due to privacy and security threats [55].

One immediate safeguard is data anonymization techniques, which unfortunately do not provide a substantial level of privacy protection to the patients while guaranteeing the performance of data analytics [56]. It is not uncommon to come across disturbing news about risks and vulnerabilities in anonymized data. For example, it is not a difficult task to "match known patients to anonymized health records in Washington state data" [57]. Netflix is under fire because of the privacy concerns and lawsuits over the anonymized database of 480,000 customers in the recommendation contest [58]. Achieving an optimal balance between model utility and data privacy is difficult when relying solely on data anonymization. Therefore, new privacy-preserving approaches are urgently needed to protect the privacy while capitalizing on the power of data analytics to build a smart and interconnected epidemic response system.

As shown in Figure 16, differential privacy provides a viable solution to address the issue of data breaches, while realizing data analytics for smart health [59, 60]. A differential-privacy algorithm ensures that one's participation in a dataset, or lack thereof, will not be disclosed [61]. Suppose that an epidemic database $D$ contains $n$ tuples, each with $d$ input variables $\mathbf{x}_i = (x_{i1}, \dots, x_{id})$ and the output space of a response variable $y_i$. The input space of $\boldsymbol{X}$ is assumed to be with symmetric neighboring relation $\mathbf{x} \cong \mathbf{x}'$. As shown in the Definition 1, privacy parameter $\epsilon \geq 0$ controls probability bounds about the level of privacy protection, which is the degree of difference allowed between output distributions by applying the function $\mathcal{F}(\cdot)$ onto databases $D$ and $D'$.

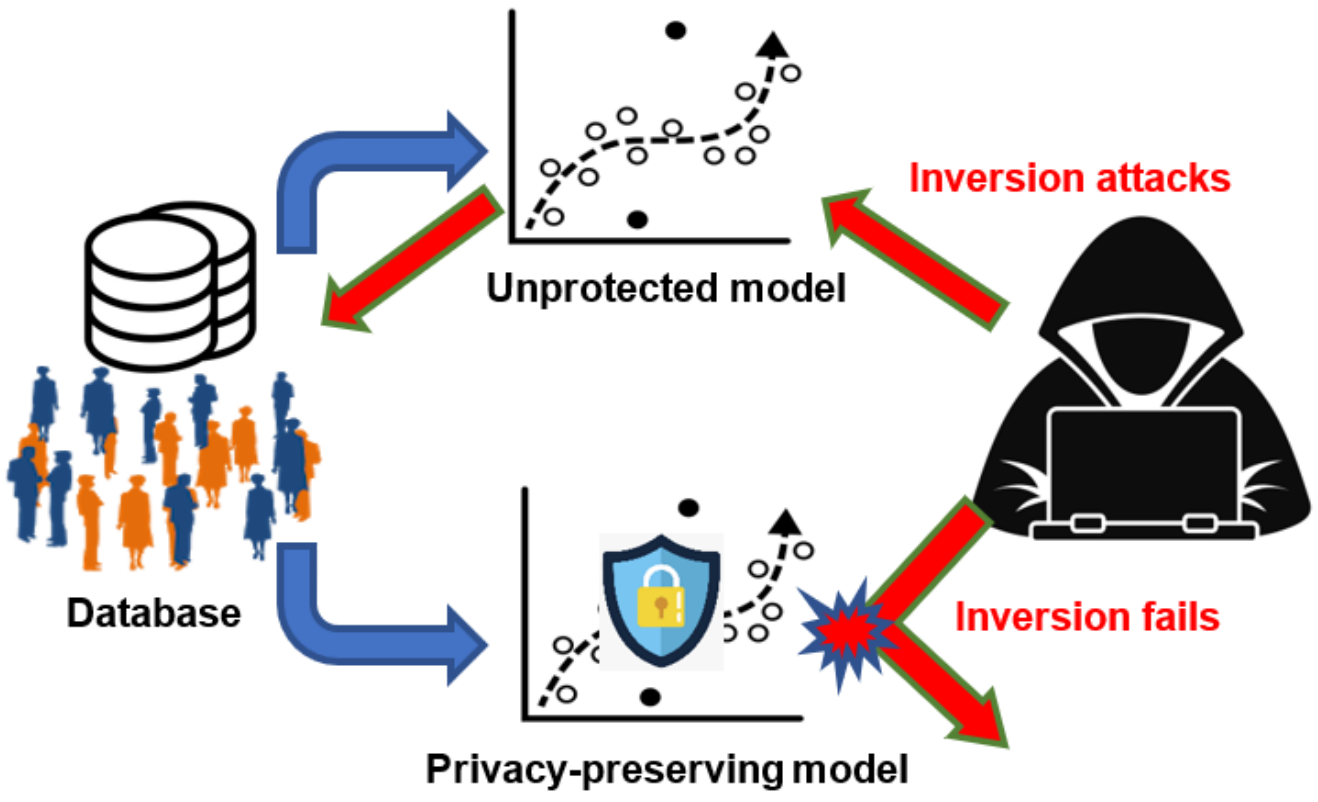

**Figure 16. An illustration of privacy-preserving predictive modeling**

**Definition 1.** *A randomized function $\mathcal{F}: X \to Y$ gives the $\epsilon$-differential privacy if for all datasets $D$ and $D'$ differing by at most one row and for all $\xi \subseteq \text{Range}(\mathcal{F})$, we have*

$$\Pr\{F(D) \in \xi\} \leq e^{\epsilon} \Pr\{F(D') \in \xi\} \quad (1)$$

Under differential privacy, one's inclusion within a dataset should make no statistical difference in an algorithm's output. Therefore, two databases that only differ by a single record of data should produce statistically similar results when running a differential-privacy algorithm [62].

The "analyze" step develops and trains machine learning models (e.g., regression models, spatiotemporal statistical models, neural networks) by minimizing the objective function $J(\beta, D)$ with the available set of predictor and response variables in the epidemic database $D = \{(\boldsymbol{x}_1, y_1), \cdots, (\boldsymbol{x}_N, y_N)\}$. Therefore, empirical risk minimization (ERM) is formulated to search an optimal set of parameters $\beta$ that minimize the regularized empirical loss function as

$$J(\beta, D) = \frac{1}{n} \sum_{(x_i, y_i) \in D} \ell(\beta, \boldsymbol{x}_i, y_i) + \Lambda R(\beta) \quad (8)$$

where $\ell$ is the loss function (e.g., prediction errors), $\Lambda$ is the regularization parameter, and $R(\cdot)$ is the regularization function. To achieve the differentially private ERM algorithms, there are three different ways to inject the designed noises (e.g., Laplacian noises) in the model training.

- **Output perturbation**: add noises to the model's optimal coefficients $\beta^* = \text{argmin}_{\beta \in \mathcal{B}} J(\beta, D)$
- **Objective perturbation**: add noises to the objective function $J(\beta, D)$
- **Gradient perturbation**: optimize $J(\beta, D)$ with noisy gradients $\nabla J$ and stochastic gradient descent

In the literature, standard output and objective perturbation techniques for logistic regression models were developed by Chaudhuri *et al.* [63]. A variant of objective perturbation, known as the functional mechanism, was later introduced by Zhang *et al.* [64]. The functional mechanism works by injecting noise into the regressor coefficients. Furthermore, the sensitive mechanism, proposed by Wang *et al.* [65], serves as an expansion to the functional mechanism. This new sensitive mechanism is capable of deterring against model inversion attacks by differentiating between sensitive and non-sensitive attributes when performing coefficient perturbation.

**Table 6. The gradient perturbation algorithm for privacy-preserving predictive modeling**

| |
|---|
| **Input:** Data $D$, parameters $\epsilon, \Lambda, \mathrm{K}, b, \theta$ |
| **Output:** Approximate noisy minimizer $\bar{\beta}$ |
| 1: Initialize $\beta^{(1)}$, $\tau = 1$, $\kappa = 1$, $\eta_0 = \sqrt{\frac{1}{\Lambda^{1/2}}}$ |
| 2: Let $\tau_0 = \frac{1}{\Lambda \eta_0}$ |
| 3: Distribute $D$ into a set of batches B, each of size $b$ |
| 4: **while** $\kappa \leq \mathrm{K}$ |
| 5: **for** each $j = 1, \dots, \lvert\mathrm{B}\rvert$ **do** |
| 6: Set $\eta^{(\tau)} = \frac{1}{\Lambda(\tau_0 + \tau - 1)}$ |
| 7: Set $\Delta^{(\tau)} = \frac{2\theta\eta^{(\tau)}}{b}$ |
| 8: Draw a vector $\mathbf{z}^{(\tau)} \sim Lap\left(\frac{\Delta^{(\tau)}}{\epsilon}\right)$ |
| 9: Set $\beta^{(\tau+1)} = \beta^{(\tau)} - \eta^{(\tau)}\left(\nabla J\left(\beta^{(\tau)}, \mathrm{B}_j\right) + \frac{1}{b}\mathbf{z}^{(\tau)}\right)$ |
| 10: Set $\tau = \tau + 1$ |
| 11: **end for** |
| 12: Set $\kappa = \kappa + 1$ |
| 13: **If** $\left\lVert\beta^{(\tau+1)} - \beta^{(\tau)}\right\rVert < \delta$, break |
| 14: **end while** |
| 15: Let $\bar{\beta} = \beta^{(\tau)}$ |

Traditionally, output and objective perturbation techniques are easy to implement and more suitable for centralized computing. Nonetheless, distributed processing has become a more dominant force in the era of big data. Gradient-based perturbation techniques provide a higher degree of flexibility in light of this distributed reality. Song *et al.* [66] first proposed the gradient perturbation for differentially private updates, which however does not adaptively adjust the learning rate for fast convergence. As shown in Table 6, this paper presents a privacy-preserving algorithm with adaptive learning rate, which is also a newly revised implementation of gradient perturbation techniques.

At the beginning of this algorithm, several parameters (i.e., a privacy parameter $\epsilon$, a regularization parameter $\Lambda$, the number of epochs K, and a batch size $b$) are firstly initialized. Note that the initial learning rate $\eta_0$ is calculated as $\sqrt{\frac{1}{\Lambda^{1/2}}}$. Next, an initial guess for regression coefficients $\beta^{(1)}$ is randomly generated. Before entering the main loop, iteration counter $\tau$ and epoch counter $\kappa$ are both set to one. The parameter, $\tau_0$, is an intermediate variable that is employed to determine $\eta^{(\tau)}$ at each iteration. The starting value of $\tau_0$ is set as $1/\Lambda\eta_0$. Further, the Dataset $D$ is divided into a set of batches B, each of size $b$.

For each epoch $\kappa$, the algorithm will process all batches B, i.e., $j = 1, \ldots, |\mathrm{B}|$. The processing of one batch constitutes a single iteration within the epoch. For each iteration $\tau$, the learning rate $\eta^{(\tau)}$ is updated, whose value is then utilized to update the global sensitivity $\Delta^{(\tau)}$. Perturbation is carried forth by drawing a random vector $\mathbf{z}^{(\tau)} \sim Lap\left(\Delta^{(\tau)}/\epsilon\right)$, which is scaled by $1/b$. This scaled noise vector is injected into the gradient $\nabla J$. Once the gradient has been perturbed, it is used to update $\beta$. The final step of each iteration entails updating the iteration counter $\tau$ by one. Once all batches are processed, the epoch counter $\kappa$ is also incremented by one. This entire process continues until convergence or until $\kappa > \mathrm{K}$.

Figure 17 shows that decreasing $\epsilon$ causes both model and attack accuracies to degrade. Nonetheless, each will decay at different rates. Once epsilon falls beneath $10^{-2}$, the attack accuracy experiences a substantial drop with minimal impact on the model accuracy. The attack accuracy approaches zero when $\epsilon$ draws closer to $10^{-4}$. However, the model accuracy only decreases by ~5% from baseline when $\epsilon$ approaches $10^{-4}$. Beyond this $\epsilon$ value, the degradation of model accuracy will accelerate. However, there is little utility in decreasing $\epsilon$ any further since the attack accuracy has already been reduced to zero. Privacy-preserving techniques provide an enabling tool to mitigate the risks and costs due to privacy breaches (e.g., model inversion attacks) while maintaining the performance of epidemic predictive models.

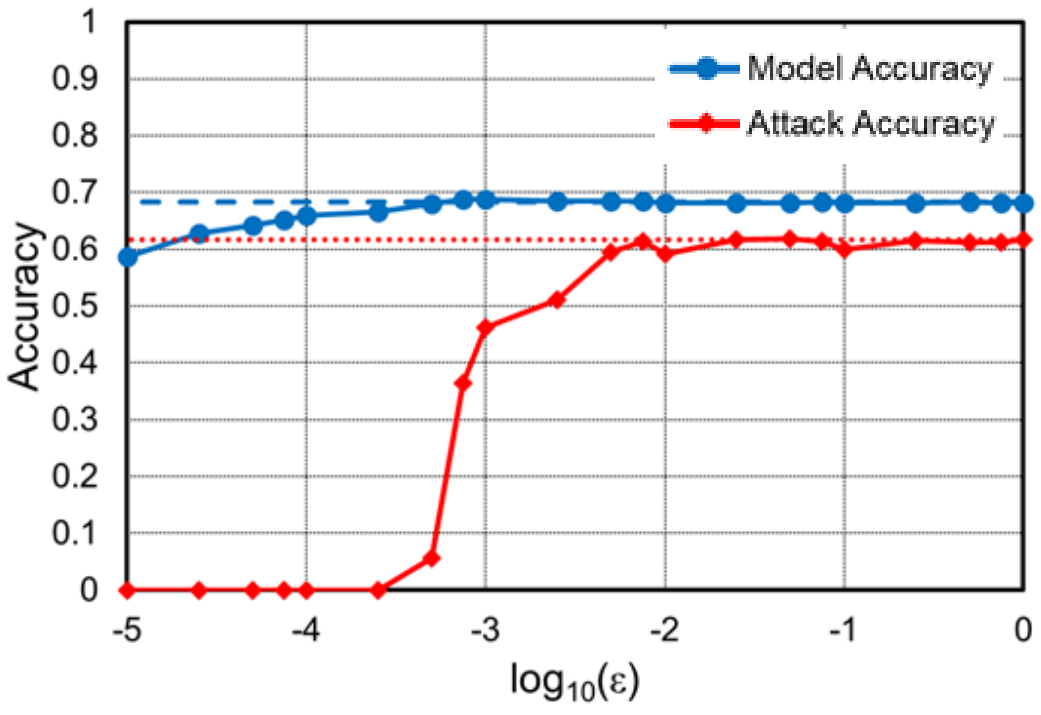

**Figure 17. Privacy model and attack accuracies with respect to varying $\epsilon$.**

## V. Improve the Resilience of Health Systems

Epidemic outbreaks demand medical resources in a short period of time. Such demands can outpace the supply for months, because the waves of an epidemic may recur in multiple years. The abrupt increase of infected cases quickly overwhelms health systems, causing the devastating shortages of staffs, beds, supplies and equipment. Also, if health systems are not prepared to handle highly contagious viruses, they will likely become "hotpots" and increase the spread of infectious diseases. The outbreak of COVID-19 urges changes and transformations of existing health systems to become a smarter and interconnected healthcare delivery system that is more resilient. The term "resilience" corresponds to the system's adaptiveness and robustness in the handling of unexpected events such as epidemic outbreaks or disasters (e.g., hurricanes, terrorist attacks, earthquakes), and can include multi-faceted definitions as follows:

- **Capacity resilience**: For increasing levels of demand, NPIs help flatten the curve to avoid the overload of health systems. On the other hand, if the curve is above capacity, a health system should be resilient to leverage the network for optimal capacity planning and allocation, as well as build up temporary capacity (e.g., field hospitals) to treat the patients and control the spread.
- **Resource resilience**: The supply chain should also be resilient to avoid shortages and provide sufficient medical resources (e.g., N95 masks, ventilators, antivirals) during epidemic events. Thus, a certain level of redundancy is needed in the design of supply chain. In addition, optimal resource allocation is urgently needed to ensure the equity and accessibility in the design of resource resilience.
- **Workflow resilience**: Traditional workflows in the hospital tend to cause secondary infections of healthcare workers and other susceptible patients. As a result, labor supply dwindles and more beds need to be allocated for the treatment of healthcare professionals. It is therefore imperative to re-design workflows and avoid secondary infections in case of an epidemic.
- **Operational resilience**: It is common that physicians see their patients in person for health care. However, with the increasing availability of wearable sensors, cloud computing, and information technology, such routine practices may be transformed to online delivery of health care or a hybrid online-onsite approach. Operational resilience calls upon the integration of telehealth systems with existing infrastructures and practices to advance the future of health care.

After all, rich data are provided in the "measure" step about the evolution of an epidemic. The "analyze" step extracts useful information from the data about epidemic characteristics. Now, the "improve" step exploits data-driven knowledge to improve the resilience design of health systems.

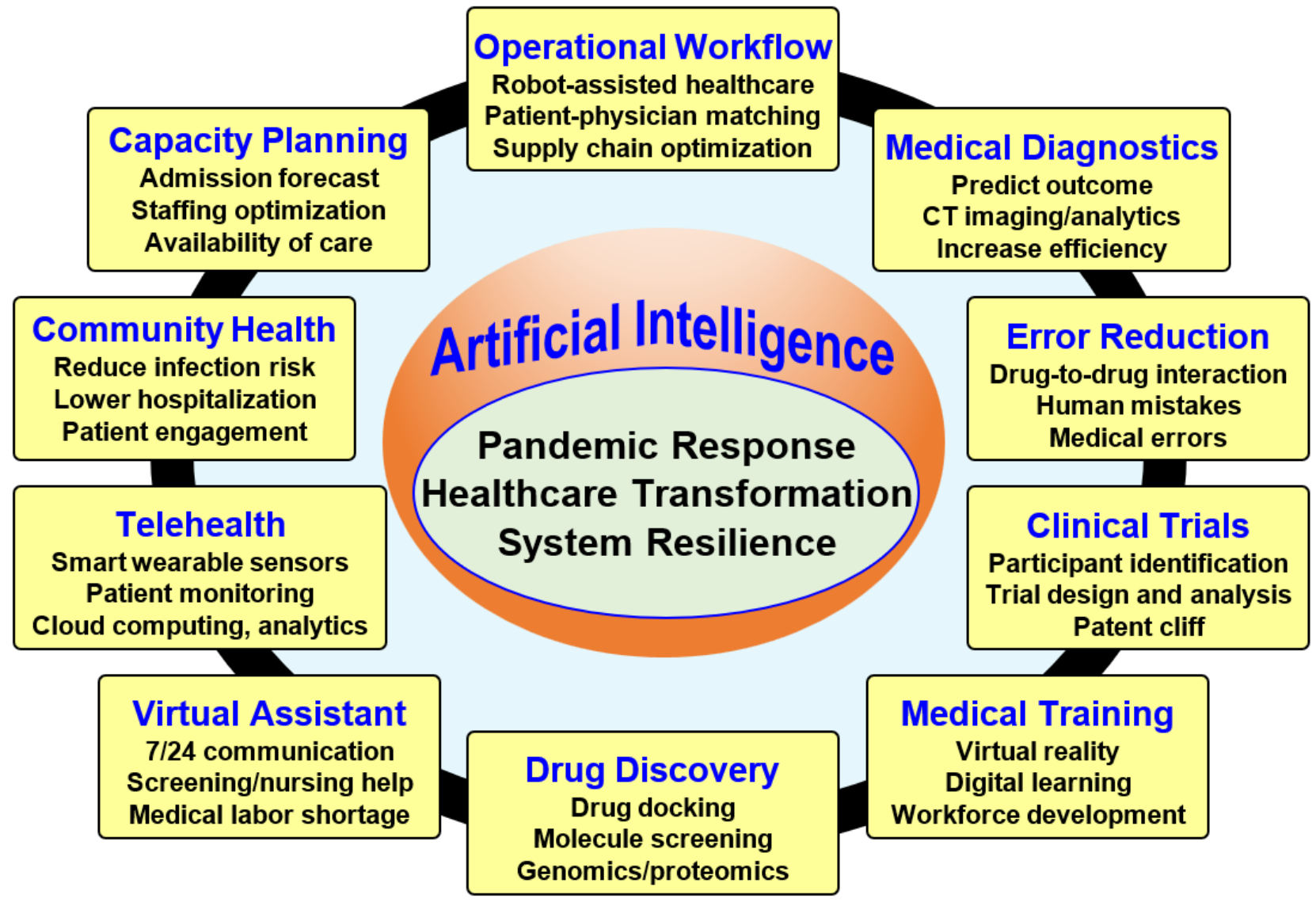

**Figure 18. AI-driven transformation of healthcare systems**

## A. Artificial Intelligence for Smart and Interconnected Health Systems

Epidemic outbreaks call upon a resilient response from health systems. As shown in Figure 18, artificial intelligence (AI) has a wide range of applications in various areas of health care, and can further promote the changes and transformation of existing healthcare practices. AI can help reduce the probability for a healthcare worker to get secondary infections via optimal allocation and use of PPEs, robot-assisted care, health informatics, and telehealth amongst many others. The provision of healthcare services includes a large number of healthcare professionals (e.g., physicians, nurses, radiologists) and medical technologies (e.g., wearable sensors, patient monitors, robotics, medical imaging). The future of work in health systems depends to a great extent on the seamless integration of human and technology. Figure 18 shows different application areas that AI has generated impacts or will bring transformations to the health systems:

- **Capacity planning**: The surge of hospitalizations is not uncommon during an epidemic outbreak. For a resilient preparation, AI tools can be developed to forecast the number of admissions, plan the medical resources, optimize the staffing level, and thereby improve availability of care. For example, the burn rate model can help health systems plan and optimize the use of PPE by predicting the trend and usage patterns during the COVID-19 pandemic [67]. Also, data-driven models can be developed to facilitate hospital planning based on the estimated demand of ICU beds, non-ICU beds, COVID admissions, and ventilators [68].
- **Operational workflow**: Infectious diseases increase the transmission risks between human subjects. AI and robots can help improve the operational workflow by human-robot collaboration, thereby reducing the probability of secondary infections. For example, autonomous service robots can take over the labor-intensive logistic tasks to deliver medications, specimens, testing results throughout the hospital 7/24 [69, 70]. AI algorithms can run the scheduling of jobs and tasks in an optimal manner. Also, operational workflow can be improved by adding an AI-supported telemedicine option, a redesign the triage process to separate infected patients from others, and smart sensing stations for symptom examination (e.g., wireless temperature sensors, thermal imaging cameras).
- **Medical diagnostics**: The increasing number of infections also brings significant amount of data in the care and treatment process. During an epidemic, a hospital is often expected to perform thousands of CT scans every day to check the status of lung infections. It is time consuming and labor intensive for human experts to visually inspect and interpret these CT images. However, AI screening is much faster than physicians, and significantly improves operational efficiency [71]. In addition, severely infected patients need advanced life support (e.g., ventilators), regular lab tests, and real-time monitoring in the ICUs. The on-hand availability of clinical data (e.g., blood pressure, heart rate, gas exchange, pulse oximetry and metabolic panel) provides an opportunity to establish AI models for the prediction of mortality and outcomes [72]. This helps stratify patients for better care and resource allocation, thereby reducing costs and the length of stay (LOS).
- **Error reduction**: To err is human, but medical errors could lead to tangible consequences. Human errors can be due to many aspects, e.g., physical condition, skill level, training, attitude, emotion and cognitive bias. AI tools can help reduce and minimize human errors, e.g., robust check of appropriate dosage levels, the reduction of diagnosis errors. For example, Liu et al. developed an AI model to offer warnings about serious side effects of drug-drug interactions [73]. This model automatically labels data from thousands of drugs and screens millions of potential interactions among drugs, which helps reduce adverse drug events and improve patient safety.
- **Clinical trials**: AI is conducive to optimize the design of clinical trials to assess and evaluate the effectiveness of drugs or antivirals for the treatment of an infectious disease. For example, AI models can be used to facilitate the identification and grouping of participants. A poor design may not reveal the effectiveness of a drug or lead to wrong conclusions about an ineffective drug. Notably, Lancet retracts a study that reported an anti-malarial drug, named *hydroxychloroquine*, has little effects to curb COVID-19 [74]. AI-based design and analysis is critical to establish the statistical significance and validity of a clinical trial. Also, pharmaceutical companies can leverage AI methods and tools to circumvent the patent cliff for the new drug design and clinical trials [75, 76].
- **Medical training**: Rapid advances of Virtual Reality (VR) technology have fueled increasing interests and steady growth in healthcare applications. For example, VR is conducive to improve medical training for decision making, and help patients to cope with pain, overcome anxiety and depression [77]. VR has been used to evaluate different kinds of medical, surgical, psychiatric, and neurocognitive conditions, as well as to improve the effects of traditional therapies in current practices [78, 79]. VR provides an immersive 3D environment for active interactions and longer training sessions. AI models can integrate sensing data with user inputs to optimize the learning steps and improve the quality of medical training [80].

- **Drug discovery**: AI has also been extensively used for drug discovery, which is evident through the rising of spotlight companies such as Genesis Therapeutics, Atomwise, and Benevolent.ai. AI is used as a preliminary filtering step to screen potential molecules on how they interact and control the activity of a virus. As there are more than billions of molecules, it is impossible for biologists or chemists to test each one of them to characterize the effects in a short period of time. AI is integrated with molecular dynamics simulation and drug docking to identify the most sensitive and effective molecules [81, 82]. Further, AI models can be developed for the analysis of genomics and proteomics, which will help gain a better epidemiological understanding of pathogen evolution and identify the origin host of a virus [83, 84].
- **Virtual assistant**: When infections and hospitalizations rise, health systems face increasing pressure due to the shortage of medical labor. There are growing concerns about the burnout and stress of physicians and nurses. Certainly, they cannot be readily available 24/7 to assist those people who are infected or are worried about getting infected by the virus. AI-embedded virtual assistants enable the patients to communicate with care providers at any time anywhere [85, 86]. The AI assistant can learn and understand the questions from a patient, screen the symptoms by steps, then guide the care.
- **Telehealth**: Telehealth provides an opportunity for healthcare professionals to deliver timely health care to patients through e-platforms (e.g., Teledoc, MeMD, iCliniq, Amwell) [87, 88]. The remote interaction greatly helps preserve the PPEs and avoid secondary infections. There are three major types of telehealth modalities as follows: (i) *Synchronous*: This is conducted through real-time live audio-video interaction with smartphone, tablet, or computer. (ii) *Asynchronous*: This includes a "store and forward" technology where message, image, or data are collected first from patients, then interpreted or responded later. (iii) *Remote patient monitoring*: This includes smart sensors and internet-of-things technologies [89, 90], where a patient's clinical data are measured and transmitted from a distance to healthcare providers. Telehealth helps screen symptoms of COVID-19, provide low-risk urgent care for non-COVID-19 patients, follow up with patients after discharge, and access to primary care or specialists for chronic disease management.
- **Community health**: AI models can be extended to the community. People live in the community and connect with social networks. Also, they commute through transportation networks. There are symptomatic and asymptomatic people who get infected, each can have a list of contacts with the risk to be infected. Contacts can be traced in the community with the use of smart sensors, mobile applications, and surveys [91, 92], e.g., Sara Alert (https://saraalert.org/). AI models can be established to characterize the spread and propagation of infectious diseases, predict the future evolution for prevention and control. In the community, AI tools can help reduce the transmission risk, prevent unplanned hospitalization, and improve the patient engagement.

AI applications are not limited to those areas describe above, and can be extended to fitness, compliance, cybersecurity, data privacy and to name a few. Epidemic response and management depend on the realization of full potentials of AI and big data to build the next-generation health system.

## B. Healthcare Resource Allocation for Coverage Control

Healthcare resources are critical to infection prevention and control during an epidemic. Examples of such resources include personal protective equipment (PPE), ventilators, medicines, antivirals, testing kits, and testing facilities (e.g., drive-thru testing sites). PPEs protect healthcare workers and patients from getting exposed to the virus. Ventilators are indispensable to saving the lives of patients with severe lung infections from coronaviruses (e.g., COVID-19) that cause excess fluid in the lungs and make patients experience difficulties to breathe on their own. Medicines and antivirals are also vital to stopping the spread of viruses and keep the mortality rate under control. Testing resources (e.g., test kits and sites) help identify infections in a timely manner and intervene as early as possible, e.g., quarantines or contact tracing to identify the patient's contacts and been-to. The availability of such resources directly determines the success or failure of virus containment before vaccines are available.

During the period of an epidemic or pandemic, the number of infected cases grows exponentially and yields heterogeneous distributions in multiple spatial regions. As such, the demand of healthcare resources is not uniformly distributed in spatial dimensions. Figure 20(a) shows the complex distribution for the number of confirmed cases in each zip code at the state of Pennsylvania. The demand varies with respect to the number of cases and population sizes across the regions, which is also called spatial demand heterogeneity. Such heterogeneity may be due to many factors, e.g., the demographic structure of the population, the infrastructure of the region, or the transportation in the area. The density of spatial demand (i.e., estimated from infected cases, demographics, or vulnerable population in each zip code) provide critical information to help optimize the allocation of healthcare resources, e.g., ventilators, testing kits, vaccines, drive-thru testing and/or vaccination sites.

The optimality, however, depends to a great extent on accessibility (e.g., shortest travel distances between demand and supply in each region) [93] and equity (e.g., the distribution of demand density among coverage regions of testing sites) [94]. Specifically, resource accessibility refers to the ease of access to resources when the demand distribution is heterogeneous in a spatial region. Equity, on the other hand, is a coverage measure of heterogeneous demands over multiple regions. A high level of equity ensures equal coverage of healthcare resources. As shown in Figure 19, let $\sigma(s)$ be the spatial demand function: $\Omega \rightarrow R^+$ that provides the demand density for each location $s$ in the polygon space $\Omega$. The objective function is to find optimal locations of resource facilities $\Theta = \{\theta_1, \theta_2, \dots \theta_i\}$ in the space $\Omega$ that minimize the sum of weighted distance functions between supply and demand locations, which is formulated as a coverage control problem. The cost function is defined as the sum of the moment of inertia in the regions:

$$C(\boldsymbol{\Theta}, \boldsymbol{V}) = \sum_{i=1}^{I} \int_{V_i} d(\|s - \theta_i\|)\, \sigma(s) ds \tag{9}$$

where $I$ is the number of resource facilities, $V_i$ is the $i^{th}$ Voronoi region, $\theta_i$ is the $i^{th}$ Voronoi center, and $\sigma(s)$ can be the demand density of confirmed cases (or vulnerable population, virus spread situations) at location $s$. The location of each resource facility is determined by minimizing the cost function as:

$$c(V_i) = \text{argmin}_{\theta_i} C(\boldsymbol{\Theta}, \mathbf{V})$$

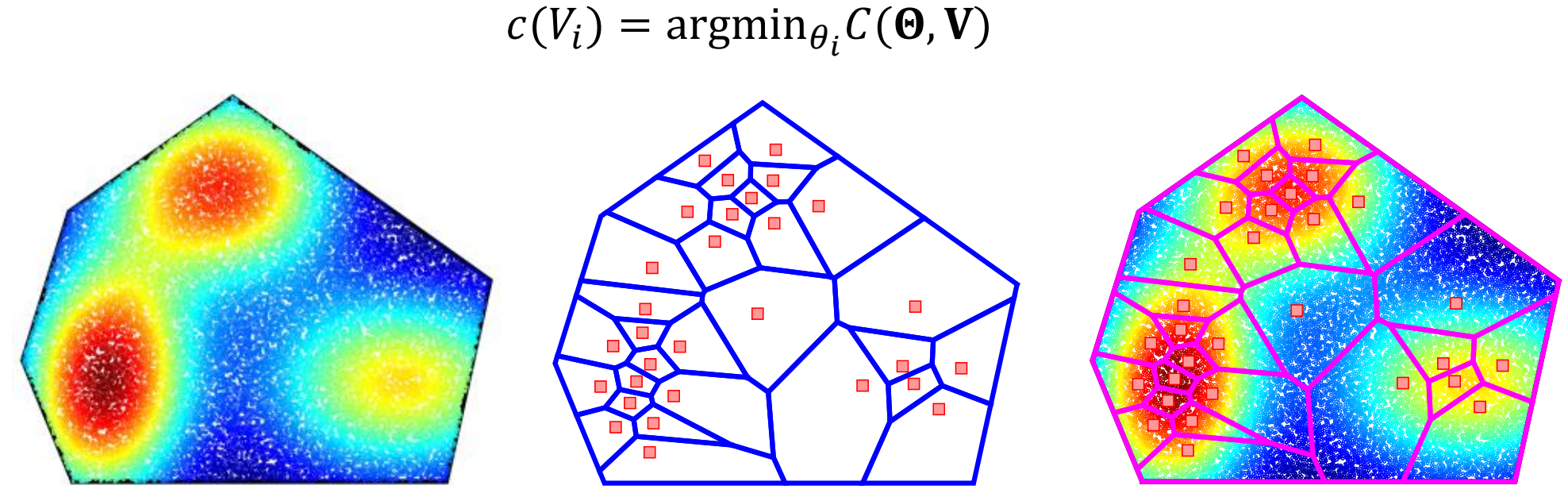

**Figure 19. Greedy-Voronoi tessellation with a heterogeneous demand function in the space**

Voronoi tessellation guarantees that each resource facility is the closest to every location in its cell [95]. The moment of inertia loss function ensures the demand density within each region is taken into account. This, in turn, helps control the coverage of heterogeneous demand over all the Voronoi regions. Note that *this is different from traditional clustering problems, because of the need to consider both distance functions and the spatial demand function that can be highly heterogenous*. Table 7 shows the proposed algorithm of greedy-Voronoi tessellation, which includes two stages, namely sequential placement and global calibration. First, each facility is sequentially placed to minimize the cost function (i.e., defined as the sum of the moment of inertia in the regions). In each iteration, Voronoi tessellation is computed based on the locations of existing facilities. A new facility is then randomly placed in the Voronoi cell with the largest mass (i.e., the sum of the moment of inertia). The tessellation is then updated to re-evaluate the new cost function. The location of this newly placed facility is optimized step-by-step along the gradient direction. This process is repeated until convergence. Such a sequential formulation provides both monotonicity and submodularity properties, and yields a sub-optimal solution. After the sequential placement, the global calibration continues to search for the optimal solution by computing and updating the Voronoi tessellation to optimize locations of all $I$ facilities. In each iteration, the locations of all facilities are adjusted and calibrated with the gradient descent algorithm. This process terminates until convergence. The algorithm then returns the optimized locations of all $I$ facilities.

**Table 7. The Greedy-Voronoi tessellation algorithm**

| Demand function $\sigma(s)$, Polygon space $\Omega$, total number of facilities $I$ | |
|---|---|
| 1: | Place the first facility $\theta_1$at the center of mass of the $\Omega$ with density $\sigma(s)$ |
| 2: | **For** $i = 2$ to $I$ |
| 3: | Randomly place a new facility $\theta_j$ in the Voronoi cell with the largest mass |
| 4: | **Compute** Voronoi tessellation $V$ based on the location of current facilities |
| 5: | **Compute** the cost function $C(\theta) = \sum_i \int_{V_i} dist(\theta_i, s)\sigma(s)ds$ |
| 6: | **Compute** the gradient $\frac{\partial C}{\partial \theta_i}$ for this newly added facility $\theta_i$ |
| 7: | Update $\theta_j$ according to $$\theta_i = \theta_i - \alpha \frac{\partial C}{\partial \theta_i}$$ |
| 8: | **Repeat** 4-7 until convergence |
| 9: | **Update** Voronoi tessellation $V$ |
| 10: | **End For** |
| 11: | **Compute** the cost function $C(\theta) = \sum_{i=1}^{I} \int_{V_i} dist(\theta_i, s)\sigma(s)ds$ |
| 12: | **Compute** the gradient $\frac{\partial C}{\partial \theta_i}$, $i = 1,2, \dots, I$ for all facility locations |
| 13: | Update all $\theta_i{}'$s according to $$\theta_i = \theta_i - \alpha \frac{\partial C}{\partial \theta_i}, i = 1,2, \dots, I$$ |
| 14: | **Update** Voronoi tessellation $V$ |
| 15: | **Repeat** 11-14 until convergence |
| 16: | **Return** facility locations $\theta_i, i = 1,2, \dots, I$ |

We implemented and evaluated the proposed algorithm using a case study of the COVID-19 infection map in Pennsylvania, as shown in Figure 20, where the red color indicates the high demand density and the blue is the low density. The proposed greedy-Voronoi tessellation helps balance between accessibility and equity for each region, allows the flexibility to dynamically adjust the tessellation based on real data (i.e., the number of available resource sites, or the variation of density in the PA map). The proposed algorithm is generally applicable to a variety of demand-driven allocation of resources in a spatial region.

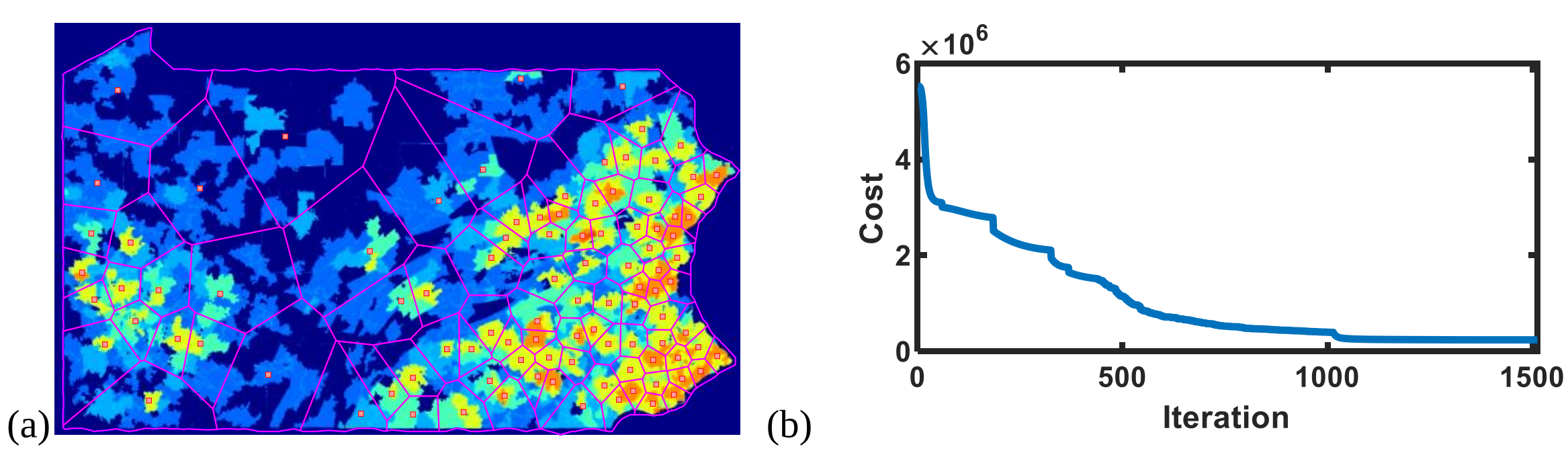

**Figure 20. (a) Greedy-Voronoi tessellation and (b) convergence curve of cost function for optimal allocation of 100 drive-thru testing sites in PA**

## C. Re-design of Health Systems

As COVID-19 cases continue to surge in the US, there is an urgent need to re-design health systems such as hospitals, medical clinics, and emergency rooms for better treatment and accommodation of patients. The main objective is to segregate infected patients, avoid secondary infections, and reduce transmission risks, thereby improving the safety and quality of healthcare services.

***Temperature screening and febrile clinics***: An immediate change is to add a screening process for patients at the entrance of a hospital, which makes it easier and less costly to detect, monitor and control the infiltration of contagious diseases [96, 97]. As shown in Figure 21, patients are required to complete a short questionnaire about their health conditions, which includes but not limited to their travel history, fever history, and symptoms. Meanwhile, their body temperature will be taken and recorded. If the answer to any of the questions on the questionnaire is "YES" or the body temperature is above the normal level, the patients will be guided to a febrile clinic, an isolated area with quarantine units at the hospital. Otherwise, they can enter the hospital/ Emergency Department (ED) and proceed to the triage area.

The identification and control of fever and high-risk patients during the screening process can separate them from other patients at an early stage, which reduces the risk of secondary infections in the hospital [98, 99]. Febrile clinics have a separate entrance and negative-pressure ventilation systems, which keep the air mix with other areas in the hospital at a minimum level. Also, febrile clinics have isolated exam rooms and its own pharmacy so that high-risk patients do not infect each other or travel to other areas of the hospital [98]. Depending on the diagnosis results, patients may either stay in the patient rooms in the febrile area to get further treatments or go home.

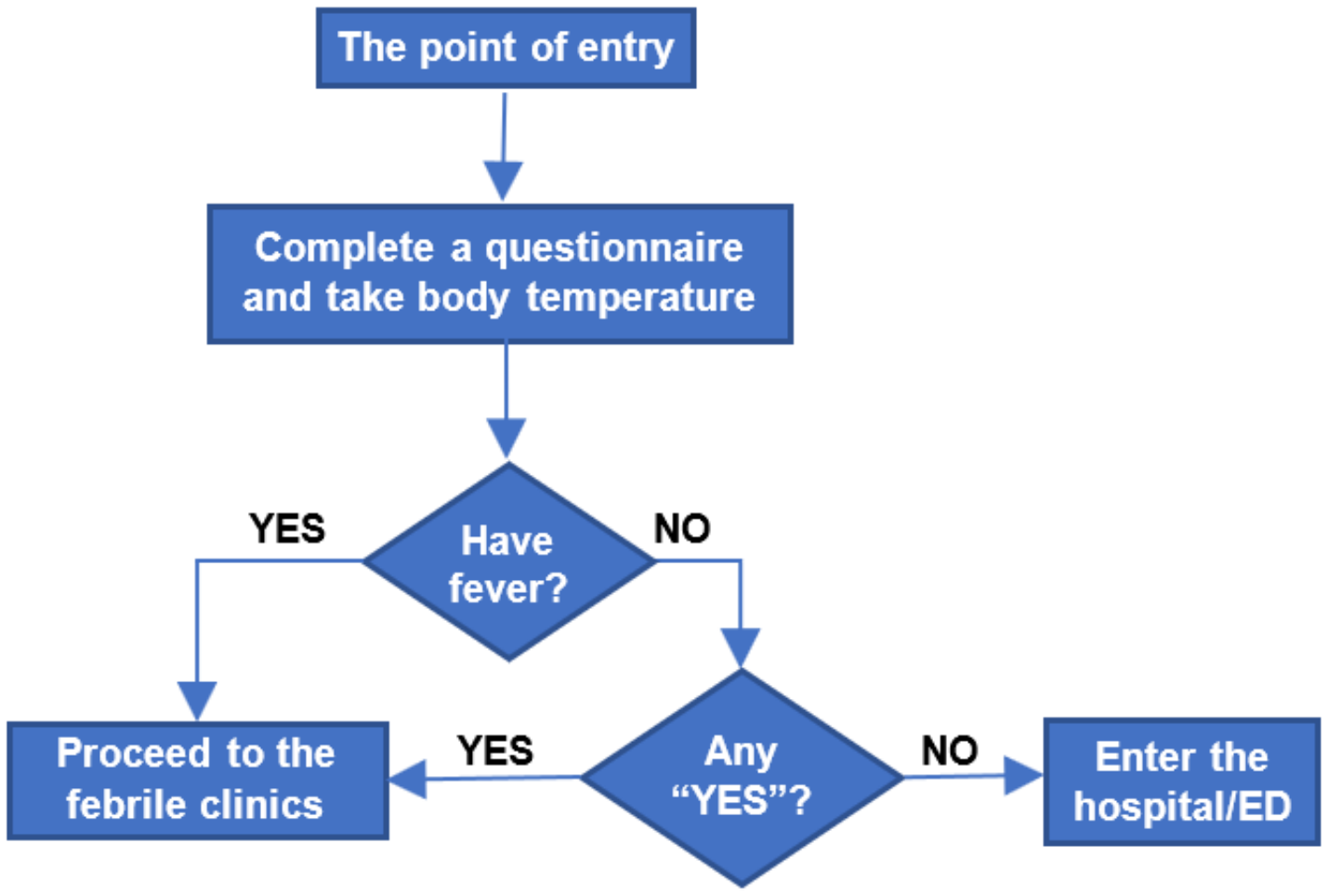

**Figure 21. The screening process and febrile clinics**

***Environmental considerations to prevent infections***: Healthcare-associated infections have caused high morbidity and mortality during the epidemic outbreaks, which are mainly due to the contact between patients and healthcare workers, patients and staff, and patients and the environment. Here, we present a brief review of recommended designs and guidelines to minimize healthcare-associated infections:

Airflow system

- The difference of air pressure between isolation rooms and other areas should be about positive 15 Pa [98].
- Room air should be changed 10-12 times every hour to sufficiently dilute the bacterial load around an infected patient [100].
- Equip ventilation, especially in communal areas [100].
- Install negative-airflow systems in areas where high-risk patients will be cared for [101].
- Isolation rooms should have negative airflow and frequent air exchanges. The air cannot be recirculated [101-103].

Hygiene and cleaning

- Install at least one sink in every patient room, examination room, procedure room and isolation room, which is close to the entrance of the room. Each sink should be with a hands-free control, soap dispenser, and paper towel holder [101, 104].
- Use information systems to monitor hand hygiene performance and provide feedback [105].
- Frequent disinfection of non-disposable material, equipment, work surfaces, wards, environment, facilities, horizontal surfaces, surfaces touched by patients and staff and toilet facilities using hypochlorite 1000 ppm [104].

Room design

- Convert the patient rooms into single rooms with en suite toilets [104, 106].
- Recommend square footage for patient rooms in critical care units (ICU): 13.94 $m^2$ per bed for single-patient rooms and 11.15 $m^2$ for multiple-patient rooms [107].
- Add ante-rooms in negative pressure rooms to reduce the escape of droplet nuclei [98].
- Equip the observation unit in the ED with isolation rooms, which have automatic doors [98].

***Drive-thru medical clinics***: As shown in Figure 22, another idea is to transform the garage of a hospital into drive-thru medical clinics. This design is currently proposed and evaluated by our University Medical Center in collaboration with NBBJ (i.e., an American global architecture, planning and design firm, http://www.nbbj.com/). Each in-car care unit will include necessary medical equipment and a paramedic. Drive-thru medical clinics bridge the gap between telehealth and in-room visits. It not only enables patients to get timely access to healthcare services, but also gain in-person interactions with healthcare providers for diagnosis and treatments. Furthermore, in-car care units provide medical monitoring that cannot be achieved otherwise via telehealth, which includes but not limited to vital signs, blood pressure, electrocardiogram, and auscultation.

In conventional medical visits, patients often go through a lengthy cycle that includes parking the car, screening in the main entrance, registration in the front desk, waiting to be called, entering the exam room, getting the services, waiting for the results, exiting the hospital, and then departing the garage. There is a higher risk for patients to get infected or infect others in each step of the process. Instead, drive-thru medical clinics follows a much shorter cycle, i.e., entering the in-car care units for examination or treatments and then departing the garage, as shown in Figure 22. For a car with four seats, four people can get health care at the same time. In-car healthcare platforms bring benefits to both patients and healthcare providers. For patients, they can get the necessary exam or treatments more conveniently and faster, while avoiding secondary infection in the hospital. For healthcare providers, they can treat more patients while maintaining the hospital capacity and have a lower risk of secondary infection.

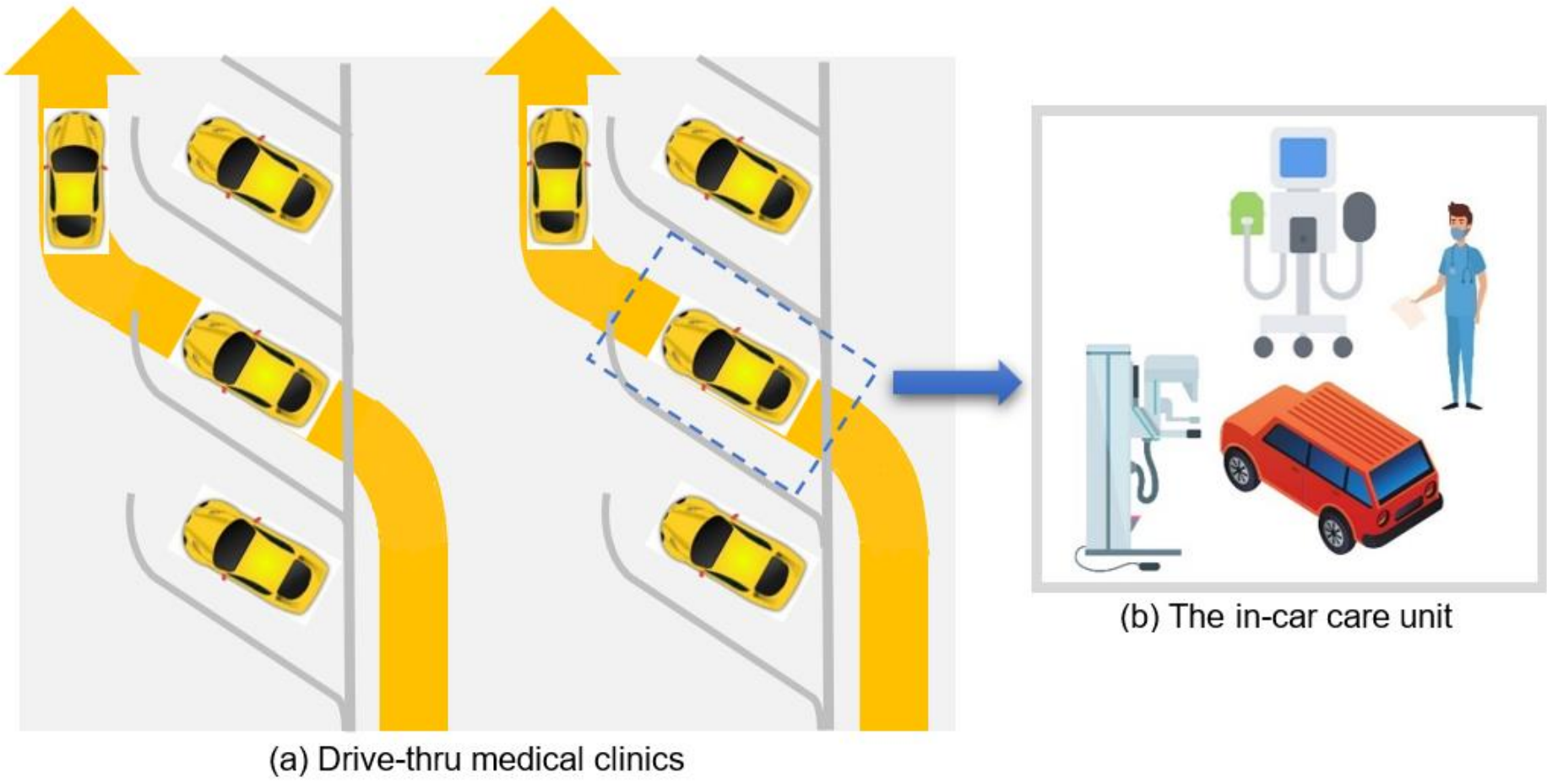

**Figure 22. An illustration of (a) drive-thru medical clinics and (b) in-car care units**

# VI. Prescriptive Analytics - Control the Spread

Predictive analytics extract useful information from the data to delineate key risk factors and predict real-time positions of virus spread. Further, this section will present the prescriptive analytics that exploit the knowledge from predictive analytics to identify the course of actions to control the spread of virus. Specifically, we will focus on the development of simulation models and computer experiments to benchmark the performances of health policies and action strategies.

## A. Simulation Modeling and Computer Experiments

Simulation models provide a mathematical description about the physics of disease propagation and how infections are correlated with the dynamics of human movements in spatial regions. With rapid advances in epidemic surveillance systems, abundant infection data are collected. The availability of data offers an unprecedented opportunity to model human traffics and the progress of an epidemic from a dynamic, as opposed to a static sense. Fast and accurate simulation models are critically needed to: (1) analyze main effects and interaction effects of process parameters in an epidemic, (2) predict how these parameters of interests impact the resource allocation and epidemic outcomes, (3) aid the design of health policies and action plans, (4) compare and benchmark a variety of existing policies and strategies, and (5) augment real-world epidemic control by providing a model-based baseline for process adjustment.

Experiments, either physical or computer-based, are critical to the discovery of new insights and knowledge from epidemic processes. However, physical experiments, also called clinical surveys or trials, on the human population are often difficult, even with the approval of an internal review board (IRB). Note that there are many practical and ethical limitations pertinent to physical experiments of human subjects. Also, it is very expensive to design a comprehensive protocol to collect data from a large population. Computer experiments with simulation models [108] are highly flexible and offers a great opportunity for the investigation of epidemic processes. As such, research communities have identified the urgent need to develop epidemic simulations and, more importantly, design computer experiments to accelerate prescriptive analytics and control the virus spread. This is essential for making the health system respond in a fast and proactive manner to disease variations and disruptive events.

From a broader vista, epidemic simulation can be categorized into two classes contingent on the level of modeling details of human behaviors, namely, continuous system dynamics modeling and discrete event simulation (DES). Continuous system dynamics models, e.g., “susceptible-infected-recovered” (or SIR) compartment models, are constructed with a set of differential equations [109, 110]. The population is assumed to be segregated into a variety of compartments (e.g., susceptible, exposed, infected, recovered), which represents different system states in an epidemic. The rates of change among these states are modeled with differential equations. Such continuous models operate at a much more aggregate level by concentrating on system states and the rates of change in sub-populations. As a result, they are more suitable to answering questions in the macro level instead of micro level. In other words, the large number of human subjects are represented as continuous states for a better description of aggregated behaviors, but individual activities cannot be tracked or modeled through the continuous system dynamics models. On the contrary, discrete-event simulation (DES) focuses more on detailed representations of individuals’ activities and environments in the spread process of infectious diseases [111]. DES models capture detailed behaviors on the individual level (e.g., movement behaviors, contact patterns, personal protective measures) and allow heterogeneity in the rates of change within sub-populations. Discrete models are generally more suitable when individual behaviors need to be modeled so that operational details are available to investigate health policies. Hence, DES models are conducive to answer specific questions in the operational or tactical level.

## B. Epidemic Simulation in the Spatial Network

Figure 23 shows our proposed DES simulation of human movements and epidemic dynamics in a spatial network. This framework is embodied by five components, namely spatial data, network modeling, human traffic, infection modeling, and computer experiments, in a close loop to investigate detailed representations of individuals' activities in spatial environments. This modeling framework is designed to overcome the complexity to model human activities directly in a spatial environment, and leverage the extracted or derived network structure to model spatiotemporal dynamics of the virus spread.

**(1) Spatial Data**: Daily activities are often inter-connected and happen in a spatial region with key activity locations such as schools, grocery stores, shopping malls, restaurants, and homes. Spatial data are readily available as geographical information system (GIS) mapping files from US Census Bureau and other geospatial service providers. Examples of mapping file formats include Environmental Systems Research Institute (ESRI) files (e.g., ArcGIS) [112], Keyhole Markup Language (KML) files (Google) [113], and shapefiles that are in a geospatial vector data format with TIGER/Line and cartographic boundary [114]. The shapefiles also include geographic entity codes (GEOIDs) that can be used to link with demographic data from the US Census Bureau. However, GIS map files contain geospatial details (e.g., forests, water wells, and rivers) that are not necessary for epidemic simulation. Although geospatial maps are static, the movement of human subjects is dynamic and tends to form traffic flows in a networked way.

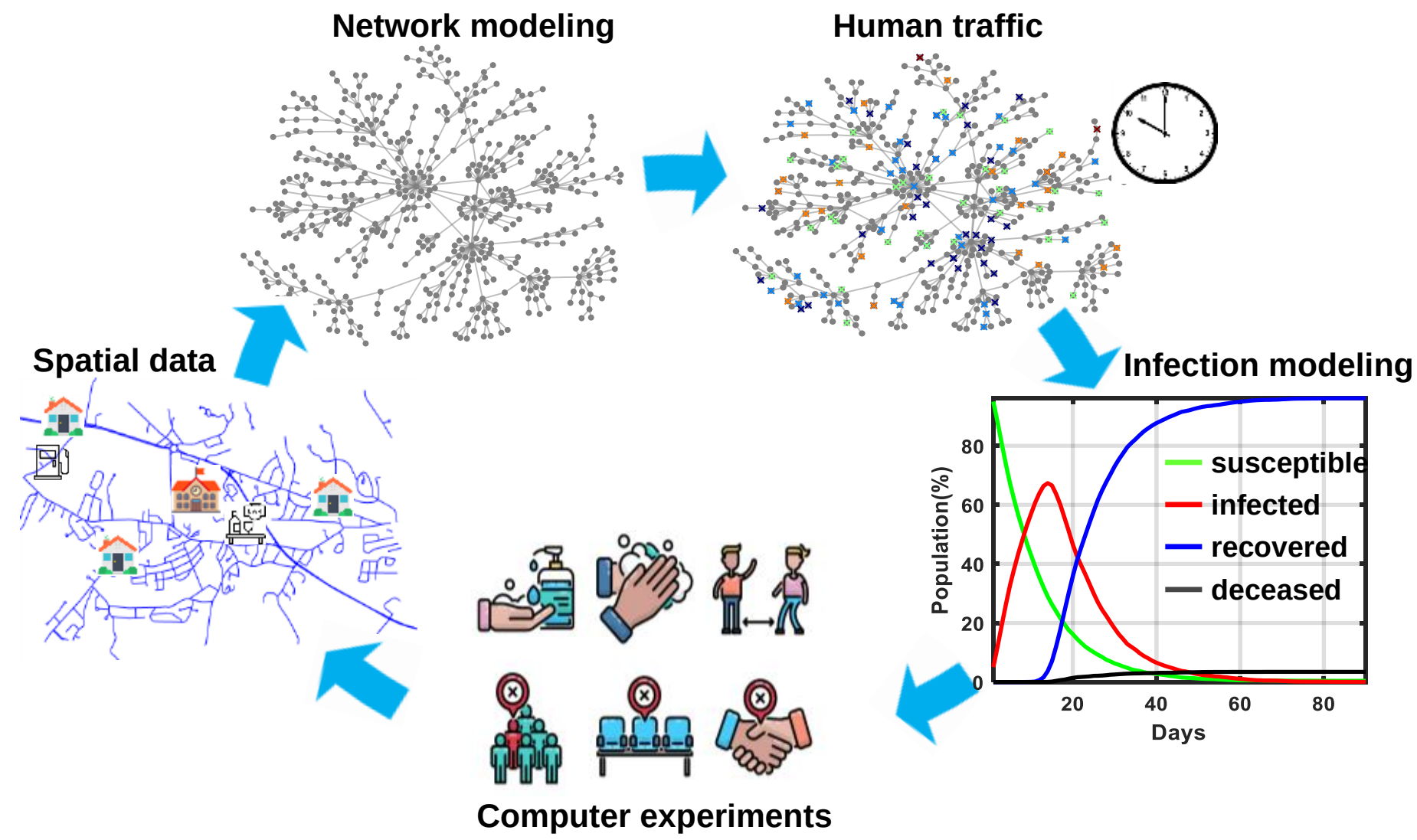

**Figure 23. The flow chart of epidemic simulation in a spatial network**

**(2) Network Model**: Indeed, many real-world systems can be represented by network models with a large number of nodes that are connected by edges or links. In a small-scale spatial environment, human subjects often visit a set of key locations (e.g., schools, stores, offices) daily that are connected by roads. In a large-scale spatial environment, people travel through a network of highways, or a network of airports. The spatial network is a graph representation with a set of nodes (i.e., key locations) that are linked by edges (i.e., spatial relationship via interconnected means of transportation). Based on real-world geospatial information, these nodes can be characterized with network features such as the degree, centrality, clustering coefficient [115-117]. In the state of the art, there are abundant literatures on network models (e.g., social network, citation network, neural network, sensor network). However, very little has been done to derive network models from GIS data in a spatial region of interest and then further investigate epidemic dynamics in the spatial network. It is imperative to model the movement dynamics of human subjects in a spatial network and further capture details of human contacts and interactions during the virus spread.

**(3) Human Traffic**: Notably, geographic entity codes (GEOIDs) can be used to link with demographic data in a spatial region. This, in turn, helps simulate the number of human subjects with a diverse set of demographic information (e.g., age groups, population sizes). The population can also be divided into different activity levels, namely low, medium and high, which correspond to the number of nodes they are going to visit. For example, individuals with a high level of activity visit more places than low and medium sub-groups in a day. These individuals will then be assigned to nodes in the spatial network, and many can be placed in the same node due to the clustered nature of residences (e.g., family members in houses and roommates in apartments). For each individual, daily activity involves the visit to a sequence of nodes via edges. The path is randomly generated according to individuals' attributes such as age groups and activity levels. We assume that individuals often choose the shortest path for each activity and therefore plan the route by Dijkstra's algorithm [118].

The schedule of human movements is simulated in a day of 24 hours as follows: The daily activity is sparse before 8 am, but become busy from 8am until midnight. The number of active individuals in the spatial network is dependent on time. New individuals will be activated and join the network traffic based on current time of a day. Rush hours are set to be at 8 am, 12 pm, and 6 pm, when more individuals will move within the network. After 11pm, no new individuals will be added, and the remaining ones will finish their activities before a new day starts.

**(4) Infection Model**: When individuals move and make contacts with each other, the virus spreads in the spatial network. The infection model provides real-time positions of healthy, infected, recovered, and deceased individuals based on human movement dynamics in the network. In this investigation, we assume that infections primarily occur in nodes, and rarely on the path. When individuals visit a sequence of nodes, they come across each other in the same node. When they share the same environment, infections occur with a certain probability by surrounded virus carriers. The infection probability is dependent on exposure time $\tau$, virus transmissibility ρ, the infectivity level $r$ of virus carriers, the number of surrounding carriers $N_r$, and the susceptibility level $s_i$ of an individual $i$. The virus transmissibility $\rho$ is a disease-specific property that defines how likely a susceptible person will be infected by the virus on average. For a virus carrier, the infectivity level $r$ defines a person's capability to infect susceptible people. In other words, some virus carriers, also called super spreaders, may have a higher infectivity level than others [119]. The susceptibility level $s_i$ defines the degree of vulnerability of an individual getting infected that may vary due to risk factors such as age, gender, and comorbidity. As such, the model of infection probability is formulated as:

$$p_i = 1 - \exp\left(\tau \sum_{r \in \langle R \rangle} N_r \ln(1 - r s_i \rho)\right), \qquad 0 < r, s_i, \rho < 1 \tag{10}$$

where $R$ is the set of virus carriers, $\langle R \rangle$ is the set of infectivity levels from surrounding carriers, and $N_r$ is the number of surrounding carriers at the infectivity level $r$. For super spreaders, $r \approx 1$.

The infected individuals can be either symptomatic or asymptomatic. For a specific infectious disease, this ratio between symptomatic and asymptomatic cases can be available when more data are collected from clinical studies [120]. For symptomatic individuals, it takes a time lag (e.g., a random variable with the mean of one day) towards self-isolation or quarantine. They will stay isolated until recovery or deceased. Asymptomatic individuals are not aware that they are a carrier of a contagious virus and will continue their daily activities. There is little time lag for a susceptible individual to get infected. However, once someone gets infected, it will take a time lag (e.g., a random variable with the mean of 14 days) to either recover or become deceased [121, 122]. Once recovered, this individual will gain an increased level of immunity to the disease. Networked traffic of human movements is integrated with infection model to study spatiotemporal dynamics of the virus spread.

## C. Computer Experiments of NPIs

The availability of simulation models enables "what-if" analysis that will help local authorities in a spatial region to dynamically adjust health policies, plan near-term health care capacity, and control virus spread with rapid and timely measures. The proposed DES simulation captures not only detailed behaviors in the individual level (e.g., movement behaviors, contact patterns, personal protective measures), but also the dynamics of population traffic for infection modeling in a spatial network. Further, we evaluate and benchmark alternative healthcare policies, akin to making informed decisions, such that the healthcare system is more resilient and can respond expeditiously and effectively to epidemic events. In the experimental setting, the spatial network contains 5000 nodes, 5000 edges, and a total number of 6000 individuals who interact with each other based on daily schedules. The total simulation time is 90 days.

**How asymptomatic vs. symptomatic impact the virus spread?** Figure 24 shows the time evolution of virus spread under three different ratios of asymptomatic vs. symptomatic cases. When the ratio decreases, there will be more symptomatic cases than asymptomatic ones. Because symptomatic cases can be quickly identified and then go into self-isolation or quarantine, the virus spreads at a slower rate. Specifically, when the ratio is reduced from 3 to 0.33, infection peaks decreases from 85.48% to 45.93%. After 90 days, the remaining susceptible populations are 0.4%, 2.18%, and 13.93% for three scenarios, respectively. At 30 days, the percentages of infected population are 81.32%, 29.03% and 2.85%, respectively. As shown in Figure 24(a-c), the percentages of recovered population at 30 days are 14.07%, 0.53% and 1.10%, and the percentages of deceased population are 0.58%, 0% and 0% for three scenarios, respectively. Therefore, temporal infection characteristics are sensitive to the variations of asymptomatic vs. symptomatic ratio.

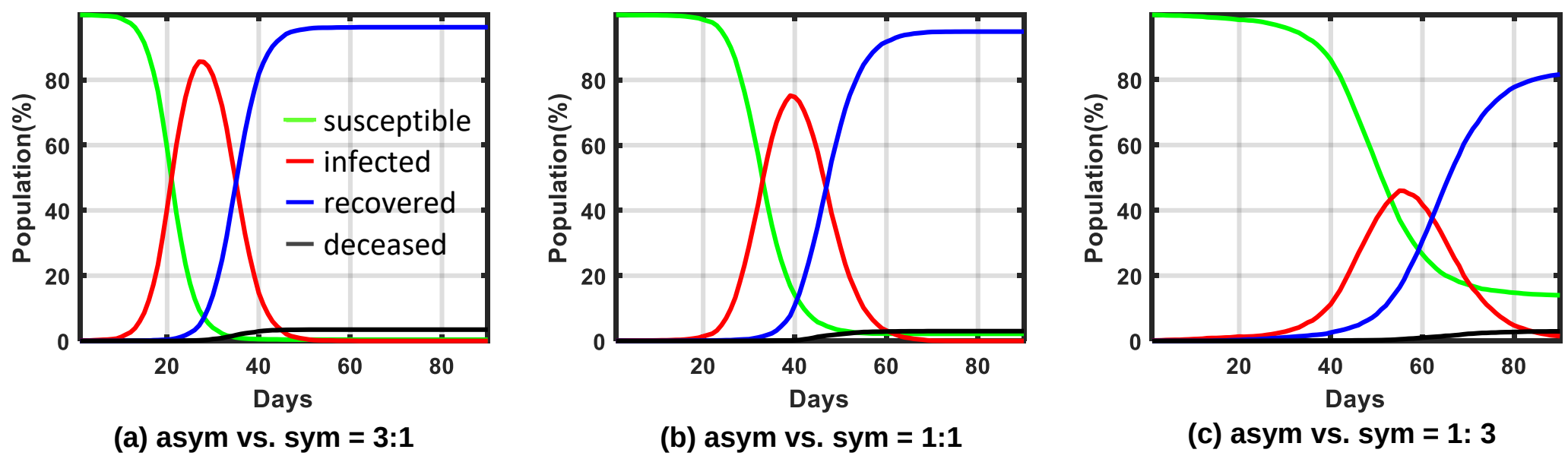

**Figure 24. Temporal characteristic curves of virus spread for different asym vs. sym ratios**

**How stay-at-home impacts the virus spread?** Figure 25 shows the impacts of the stay-at-home policy on the time evolution of virus spread with three different transmissibility values (i.e., $\rho$=0.00025, 0.0005 and 0.001). The solid line represents the implementation of stay-at-home policy that reduce daily activities to 67%, while the dashed line represents the scenario with regular activities. After the stay-at-home policy is enforced, Figure 25 (a-c) shows that infection peaks drop dramatically from 49.83%, 85.48%% and 98.20% to 2.3%, 41.60% and 80.52%, respectively. When daily activities are reduced, the time to reach infection peak is also prolonged. This time delay decreases when the virus transmissibility increases. Hence, the stay-at-home policy is critical to stopping the virus spread and flattening the curve, which will provide tremendous help to avoid an overload on the healthcare systems.

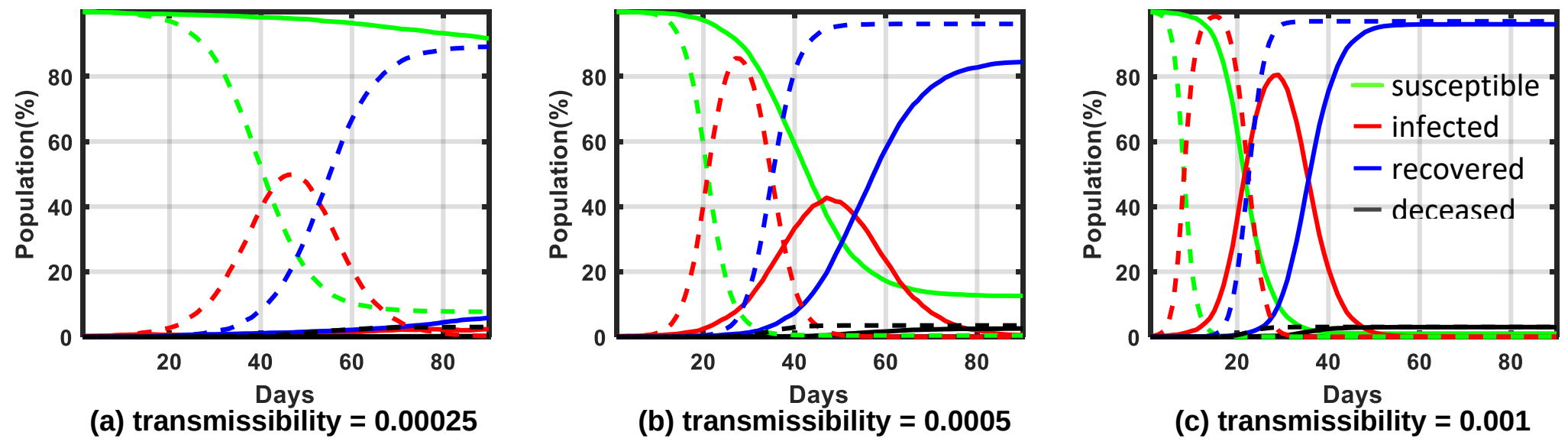

**Figure 25. Temporal characteristic curves of virus spread under stay at home policy**

**How non-pharmaceutical interventions impact the virus spread?** Figure 26 shows the impacts of NPIs on the virus spread that is benchmarked with the baseline scenario (i.e., regular daily activities without interventions, and the ratio of asym vs. sym is 3:1). When infections exceed 20% of the population, the stay-at-home policy is triggered to reduce the level of daily activity to 67%. As shown in Figure 26(b), this intervention decreases the increasing rate of infections (i.e., the derivative of the red line), and the infection peak is much lower than the baseline scenario. Further, protective measures are triggered for active individuals when more than 30% of the population gets infected (e.g., good hygiene, face masks, social distancing). Figure 26(c) shows that this policy greatly reduces the increasing rate. Also, when the disease transmissibility decreases with protective measures, the proportion of susceptible population after 90 days increases from 7.10% to 30.00%. Eventually, the proportion of deceased population is approximately 3%, 2.2% and 1.2%, respectively, as shown in Figure 26(a-c). NPIs reduce the transmission risks of infectious diseases. Therefore, a combination of NPI policies should be implemented to effectively lower the probability of infection and save more lives.

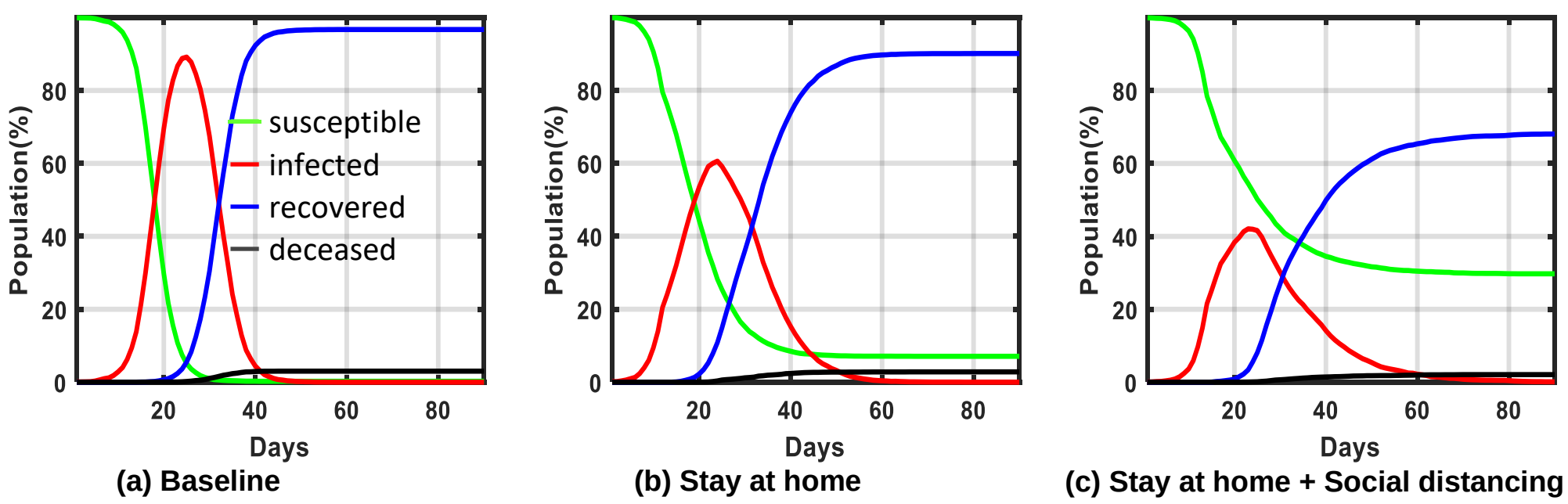

**Figure 26. Temporal characteristic curves of virus spread under NPIs**

Simulation-based decision support provides an enabling tool to benchmark alternative healthcare policies and make health systems more resilient to coronavirus events, rather than relying solely on the experience and expertise of human experts. The proposed DES simulation provides detailed behaviors of individuals (e.g., movement behaviors, contact patterns, personal protective measures) in a spatial network. Such details are often not available in conventional DES, SEIR, or statistical models, and therefore can be used to help design and analyze clinical testing programs for the population in the future work. Furthermore, networked traffic of human movements offers a higher level of flexibility for future investigation of network interdiction models in the epidemic settings. In other words, public health experts will be able to investigate the traffic control through arc interdictions to stop the spread of infectious diseases. In summary, effective simulation analysis and prediction of virus positions in geographic regions will not only help optimize the design of healthcare policies to control the virus spread, but also help safeguard the population and make health systems more resilient to epidemic events. The proposed methodology can be applicable in general to a wide range of infectious diseases.

# VII. Conclusions

The broad spread of a highly infectious disease leads to an epidemic in a country and may also bring a global pandemic if ravaging over multiple countries. For example, COVID-19 changes everyone's daily life and poses significant challenges to health and economy of our society. Before vaccines or antivirals are available, non-pharmaceutical interventions (e.g., isolation, quarantine, hygiene, face masks and social distancing) are only effective means for the control and containment of virus spread. This does not change much in the 21st century, although health systems are equipped with more advanced technologies than the era of 1918 Spanish flu epidemic. However, modern health systems do have the increasing capability of medical testing and diagnostics for a specific virus, with rapid advances of gene/DNA, microbiology, and imaging technologies. As such, large amounts of data are collected in the evolving process of epidemic outbreaks. The availability of data calls upon the development of new analytical methods and tools to gain a better understanding of virus spreading dynamics, optimize the design of healthcare policies for epidemic control, and improve the resilience of health systems.

This paper presents a holistic approach about epidemic informatics and control in the framework of **D**efine**, M**easure, **A**nalyze, **I**mprove, and **C**ontrol (DMAIC), which focuses on the intensive use of data, statistics and optimization. The proposed DMAIC framework integrates epidemic data with statistics, AI, privacy, system design, and simulation models to predict real-time positions of virus spread in the spatial network, simulate human traffic and virus spread dynamics, and provide decision support tools for the design of healthcare policies. As opposed to purely data-driven approaches, which cannot suggest action strategies, this DMAIC framework provides a higher level of flexibility to not only design computer experiments for the analysis of a variety of alternative health policies and strategies, but also augment real-world epidemic control by providing a model-based baseline for process adjustment. In addition, epidemic surges mandate the re-design of health systems such as hospitals, medical clinics, and emergency rooms for better treatment and accommodation of patients. Such re-designs help segregate infected patients, avoid secondary infections, and reduce transmission risks, thereby improving the safety and quality of healthcare services. System informatics show strong potentials to spur the growth of healthcare innovations in the US and the world, as well as complement the pharmaceutical and medical approaches to stop the spread. We hope this review can help catalyze more in-depth investigations and multi-disciplinary research efforts to advance the system informatics methods and tools for the future of healthcare.

## Acknowledgement

The authors of this work would like to acknowledge NSF Rapid grant IIP-2026875, NSF I/UCRC Center for Healthcare Organization Transformation (CHOT), NSF I/UCRC award #1624727, and Penn State Health for supporting this research. Any opinions, findings, or conclusions found in this paper are those of the authors and do not necessarily reflect the views of the sponsors.